\documentstyle[ltwol,float]{article}

\input epsf.tex

\def\Journal#1#2#3#4{{#1} {\bf #2}, #3 (#4)}

\def\NPB{{\em Nucl. Phys.} B}
\def\PLB{{\em Phys. Lett.}  B}
\def\PRL{\em Phys. Rev. Lett.}
\def\PRD{{\em Phys. Rev.} D}
\def\ZPC{{\em Z. Phys.} C}

\def\kt{$k_T$}
\def\avkt{$\langle k_T \rangle$}
\def\pt{$p_T$}
\def\s{$\sqrt{s}$}
\def\et{$E_T$}

\def\etmax{$E_{Tmax}$}
\def\x{$x$}

\def\be{\begin{equation}}
\def\ee{\end{equation}}
\def\bea{\begin{eqnarray}}
\def\eea{\end{eqnarray}}

\input psfig

\begin{document}

\title{QCD at High Energy (hadron-hadron, lepton-hadron, gamma-hadron)}

\author{J. Huston}

\address{Physics and Astronomy Dept., Michigan State University, 
East Lansing, MI 48824 USA\\E-mail: huston@pa.msu.edu}


\twocolumn[\maketitle\abstracts{ This talk summarizes recent QCD results
from HERA, the Tevatron Collider and Tevatron fixed target experiments.}]

\section{Introduction}

	As implied by the title of this talk, the topics to be discussed
cover a very wide range, encompassing QCD results from the Tevatron 
$\overline{p}p$ collider, the HERA $ep$ collider and the Tevatron fixed
target experiments. In my talk, I will not try for a totally comprehensive
review, but instead will discuss some of the important experimental and
phenomenological developments in perturbative QCD since the last Rochester
conference. 
I will not be covering deep inelastic scattering
results per se, which will be discussed in the summary talk of Tony Doyle;
instead I will present results which 
emphasize the details of the hadronic final state in DIS
and photoproduction events. Similarly, I will not discuss diffraction which will
be the subject of the review talk by Martin Erdmann. Many of the recent 
developments in the theory/phenomenology world, along with a discusion
of our current
understanding of ${\alpha}_s$, will be contained in the talk of Yuri
Dokshitzer.

	The main theme of my talk will be the success with which perturbative
QCD has been applied to the data from Fermilab and from HERA. There are enough
mysteries left, however,  
to make life interesting (and to provoke the need for larger
data samples), with several of the mysteries involving the remaining 
uncertainties in the gluon distribution. DGLAP-based perturbative QCD 
predictions remain very successful and the search for convincing evidence
for BFKL effects continues.

\section{Tevatron Collider}

In the Tevatron collider, 900 GeV protons collide with 900 GeV antiprotons
leading to a center-of-mass energy of 1.8 TeV, the highest energy
currently accessible.
The Tevatron collider completed a very successful Run 1 in 1996 with
each experiment (CDF and D0) accumulating on the order of 100 $pb^{-1}$ of 
data. Most analyses have been published or are nearing publication. 

\subsection{Inclusive Jet Production at the Tevatron}

	The inclusive jet cross section in the central rapidity region has
 been measured by
both the CDF and D0 experiments at a center of mass energy of 1.8 TeV. 
Jets are defined
using an iterative fixed cone algorithm with a radius R 
($\sqrt{{\Delta}{\eta}^2+{\Delta}{\phi}^2}$) 
of 0.7.~\cite{jetcone}
The measurement spans the transverse energy range from 15 GeV/c to the order
 of 500
GeV/c; in this range the jet cross section drops by over 9 orders of magnitude.
The highest \et\ jet events probe the smallest distance scales ($10^{-17}$ cm) 
currently  accessible. Any new physics that might exist at these distance 
scales,
such as compositeness, might
manifest itself in the jet cross section measurement.

	The jet cross sections from both experiments 
 are corrected for detector measurement and resolution 
effects and are compared to next-to-leading order (NLO) QCD 
calculations.~\cite{aversa,jetrad,eks} The 
theoretical uncertainties in inclusive jet production are greatly reduced at
NLO as 
compared to leading order. The two programs that are currently in use are 
JETRAD~\cite{jetrad}
and EKS.~\cite{eks} JETRAD generates the NLO inclusive jet cross section
by a  Monte Carlo phase space slicing technique
 while EKS is an analytical calculation. Both
programs are
implementations of the same matrix elements~\cite{ellis} and give essentially
equivalent 
results when the same cuts/conditions are applied. At NLO, the sensitivity of
the jet
cross section to the renormalization/factorization scale is reduced, but
still present. 
The value of this scale should be proportional to the hardness of the scatter.
It is convenient to set the renormalization and factorization scale equal to
a multiple of the $E_T$ of the measured jet. One can also set these scales to
a multiple of the maximum of all jets in the event ($E_{Tmax}$), at the cost
of introducing another variable into the prescription. 
Typically,
$E_{Tjet}/2$ is used in the EKS program while $E_{Tmax}/2$ is used in 
JETRAD.~\footnote{In the JETRAD program, each Monte Carlo event has 2 or 3
partons in the final state, leading to a possibility of either 2 or 3 jets.
The $E_T$ of each individual jet is not known until a jet clustering algorithm
has been applied (the two lower $E_T$ partons may be clustered if they are 
close together). The only scale known unambiguously at the time of the event
generation is the $E_T$ of the most energetic parton ($E_{Tmax}$). A version of
JETRAD also exists in which a second pass is made through the generated 
events, allowing the use of the scale $E_{Tjet}$.} 
$E_{Tjet}/2$ may be a 
more ``natural'' choice for an inclusive jet cross section, but \etmax\ /2
is also acceptable.
The use of $E_{Tmax}/2$ rather than $E_{Tjet}/2$ leads to a reduction
in the jet cross
section of
~7\% at \et\ = 50 GeV/c decreasing to $<1$\% at \et\ = 100 GeV/c. 
The effect on the NLO
inclusive jet cross section of variations in the renormalization/factorization
scale, the
value of $R_{sep}$ (the minimum separation of the two partons for them to be
considered
as two separate jets), and the choice of parton distribution functions
(pdf's) is 
investigated in more detail in Ref. 6.

\begin{figure}[t]
\center
\psfig{figure=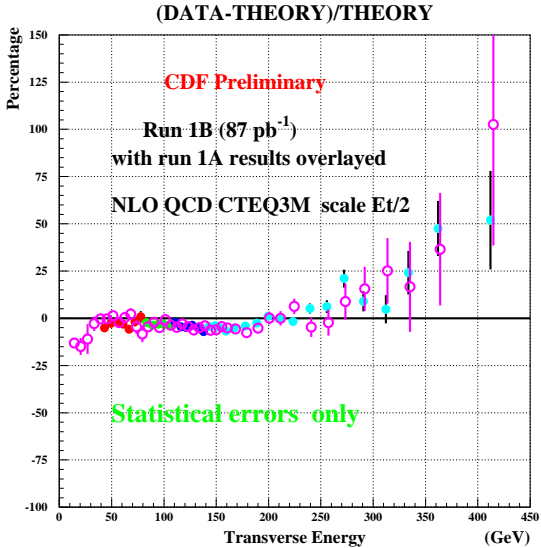,height=3.5in}
\caption{The inclusive jet cross section from CDF for Run 1A and Run 1B
compared to the NLO QCD prediction using the CTEQ3M parton distributions.}
\label{fig:cdf1a1b}
\end{figure}

	The picture that has been adopted for jet production is that the
final state
consists of 2 or 3 partons from the hard scatter accompanied by an underlying 
event due to the collision of the proton and antiproton remnants. The underlying
event is taken to be identical to that observed in minimum bias events and its 
contribution to the jet cone energy is subtracted before any comparisons to
theory. This picture has been successful but may be incomplete; there may be
additional contributions to the underlying event from double parton scattering
and higher order radiation effects that may not be included in the subtracted
minimum
bias level and may not be correctly described by the NLO
QCD calculations.~\cite{pumplin,webber,xtexplain}
A uniform extra contribution to the jet energy might manifest itself as a jet
profile broader in experiment than in theory. Such an effect has already been
observed by both CDF and D0. The main impact of any
underestimate of the underlying energy level in jet events would be on lower
energy jets.  

	The CDF collaboration has previously published the inclusive jet 
cross section
from Run 1A (19.1 $pb^{-1}$) for $0.1<|{\eta}|<0.7$.~\cite{cdfjet1a} 
A linear comparison of 
(Data-Theory)/Theory is shown in Figure~\ref{fig:cdf1a1b} 
along with the preliminary results from
Run 1B(87 $pb^{-1}$).  Good agreement with the NLO prediction is observed except at
the highest values of transverse energy. The excess is inconsistent with the
(highly
correlated) systematic error of the measurement and cannot be explained by a 
different  choice of renormalization and/or factorization scale, or by a 
different choice of
conventional parton distribution function.

\begin{figure}[t]
\center
\psfig{figure=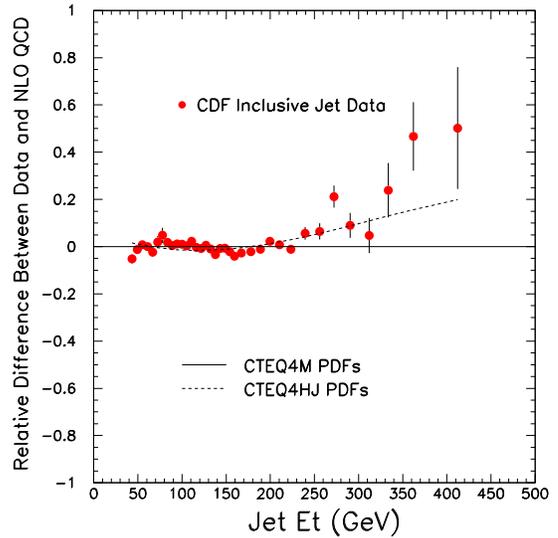,height=3.5in}
\caption{The CDF inclusive jet cross section from Run 1B compared to NLO QCD
predictions using the CTEQ4M and CTEQ4HJ parton distributions.}
\label{fig:cdf4hj}
\end{figure}

	The CTEQ collaboration has performed a global pdf fit using the 
Run 1A jet data
from CDF,  giving a large emphasis in  the fit to the high \et\  data. The
resulting fit
(CTEQ4HJ) contains a gluon distribution substantially greater (by a factor
of 2 at x=0.5)
than that in  conventional pdf's.~\cite{cteq4hj} 
The larger gluon distribution at high 
x leads to a greater
cross section at high \et\ (20\% at 450 GeV/c). The increased jet cross section 
prediction does not
pass directly through the center of the high \et\ data points, but does pass 
through the 
bottom of the error bars. The CDF jet cross section from Run 1B is shown in 
Figure~\ref{fig:cdf4hj} 
compared to the NLO QCD predictions using the CTEQ4M and CTEQ4HJ 
pdf's. 

	For high \et\ jet production, the dominant subprocess is  
$\overline{q}q$ scattering. 
The gq subprocess comprises only ~20\% of the cross section and the gg
subprocess contribution is minimal. For this reason, 
an increase
in the gluon distribution of a factor of 2 leads to only a 20\% increase in the 
jet cross section. 
The quark distributions are also free to change in this fit, but are 
tightly constrained by
the highly precise deep inelastic scattering (DIS) and Drell-Yan (DY) 
data at these x values. 

\begin{figure}[t]
\center
\psfig{figure=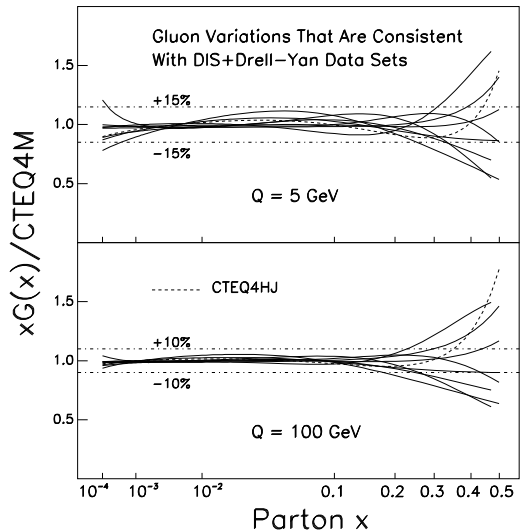,height=3.5in}
\caption{The ratio of gluon distributions consistent with the DIS and DY
data sets to the gluon distribution from CTEQ4M. The gluon distribution from
CTEQ4HJ is also shown for comparison.}
\label{fig:gluonuncert}
\end{figure}

	It may seem suprising that the gluon distribution has this degree of 
flexibility. A
recent CTEQ paper explored the uncertainty in the gluon distribution by 
performing
a gluon parameter scan, utilizing the DIS and DY data used in the 
CTEQ4 fit.~\cite{cteqgluon} The 
resulting pdf's were excluded if there were any clear conflicts with any of
the data sets.
The pdf's that remain are shown in Figure~\ref{fig:gluonuncert}
and indicate that the gluon is 
tightly constained
at lower x.  DIS and DY data provide little constraint, however,
on the high x gluon  distribution and this is demonstrated in the much
wider variation observed in the large x region. (The CTEQ4HJ gluon 
distribution is also indicated for comparison purposes.) 
In previous pdf's this constraint has been provided either by fixed target 
direct photon 
data~\cite{pdfphot} and/or jet data from the Tevatron collider.~\cite{CTEQ4M} 
Due to evolution effects, the gluon distribution is more tightly constrained
(for $x<0.2$) at high $Q^2$ than at low $Q^2$. 

\begin{figure}
\center
\psfig{figure=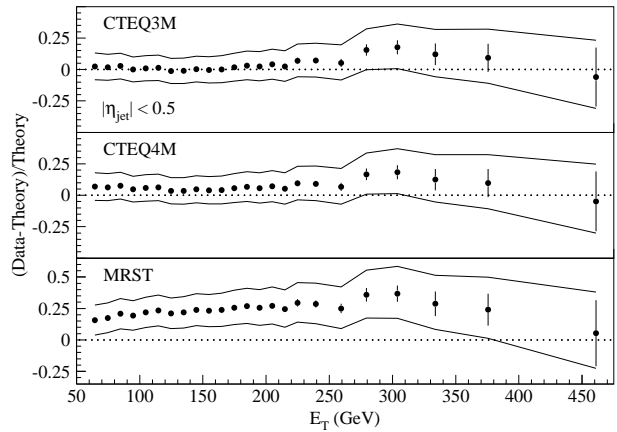,height=2.25in}
\caption{A comparison of the D0 inclusive jet cross section and NLO  QCD
(JETRAD)
predictions obtained using three different parton distribution sets. The band
represents the total experimental uncertainty.}
\label{fig:d0fig3}
\end{figure}

	D0 has presented at this conference  a measurement of the 
inclusive jet
cross section for $|\eta|<0.5$ (and $0.1<|\eta|<0.7$ 
for direct comparison to CDF) with a 
substantially reduced systematic error.~\cite{blazey,470} 
The cross section for $|\eta|<0.5$ is 
compared to 
NLO QCD predictions with several pdf's in Figure~\ref{fig:d0fig3}. 
Good agreement is 
observed with
perhaps some sign of an excess at moderate to high \et\ when
the CTEQ3M and CTEQ4M 
pdf's are used. 
The data is uniformly larger than the prediction using the MRST pdf. 
This latter deviation 
is due to the substantially weaker MRST gluon distribution at high x 
(see Section 3.1). 

\begin{figure}
\center
\psfig{figure=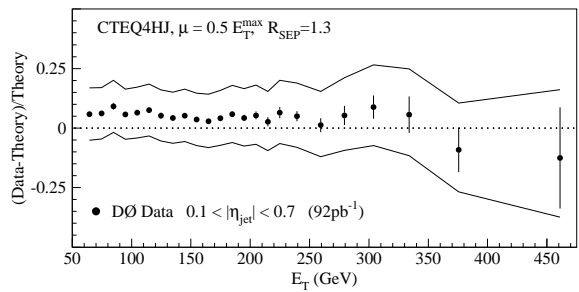,height=1.5in}
\caption{A comparison of the D0 inclusive jet cross section and a NLO 
QCD (JETRAD)
prediction using the CTEQ4HJ parton distribution set.}
\label{fig:d04hj}
\end{figure}

	Taking into account the correlations in the systematic errors, 
reasonable $\chi^2$
values are obtained for all 3 pdf's. 
The MRST prediction agrees in shape 
with the D0 data, although not in normalization. 
The best $\chi^ 2$ agreement 
(nominally better than
MRST) is obtained with the CTEQ4HJ pdf (shown in Figure~\ref{fig:d04hj}
for $0.1<|\eta|<0.7$). 

	A reanalysis of data from SLAC and NMC, taking into account nuclear
binding effects in the deuteron, predicts a larger d quark distribution at 
high x than found in modern pdf's.~\cite{unki} One of the consequences
of this larger high x d quark distribution would be an enhanced high $E_T$
jet cross section (by about 10\% at the highest $E_T$ values). 

\begin{figure}
\center
\psfig{figure=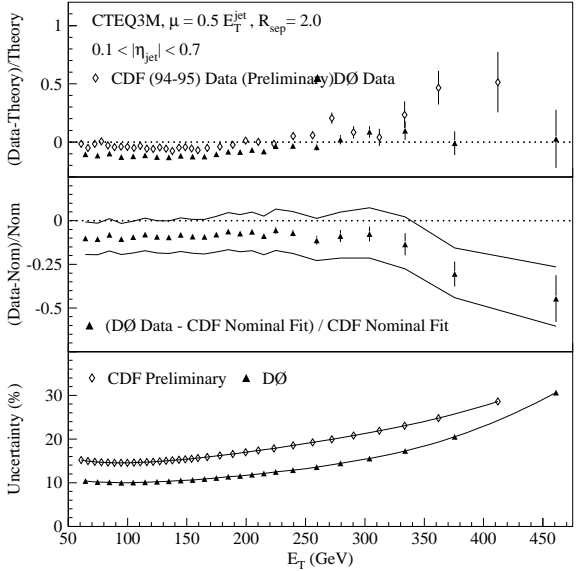,height=3.0in}
\caption{The top plot shows the normalized comparison of the D0 inclusive
jet cross section to NLO QCD predictions (EKS) along with a similar comparison
of the Run 1A CDF inclusive jet cross section. 
The CTEQ3M parton distribution set is 
used. The middle plot indicates the difference between the D0 inclusive jet
cross section and a smooth fit through the Run 1A CDF inclusive jet cross 
section, normalized to the latter. The band represents the uncertainty on
the D0 data. A comparison of the CDF and D0 jet systematic errors is shown
at the bottom.}
\label{fig:cdfd0jet}
\end{figure}

	A direct comparison of the CDF Run 1A and D0 jet cross sections is 
shown in Figure~\ref{fig:cdfd0jet}.
Aside from a normalization shift of ~5\%, the two experiments obtain very 
similar jet
cross sections, with the most noticeable difference being in the last two
data points. 
 A normalization shift of this magnitude is to be expected 
since CDF 
and D0 use different values for the total inelastic $\overline{p}p$
cross section. D0 has calculated a low probability for their jet data to
agree with the physics curve drawn through the CDF Run IA data, but if the
correlated errors for both experiments are taken into account, there is 
agreement between the two at the 46\% level of probability. 

\subsection{Dijet Cross Sections at the Tevatron}

\begin{figure}[t]
\center
\psfig{figure=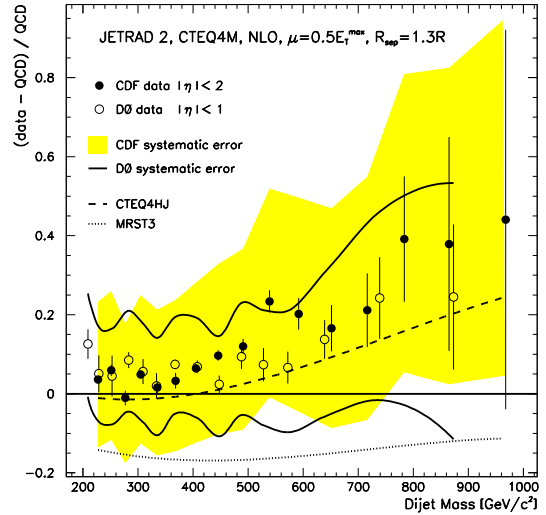,height=3.5in}
\caption{A comparison of the CDF(preliminary) and D0~$^{19}$
 dijet mass cross sections to the 
NLO QCD predictions using the CTEQ4M, CTEQ4HJ and MRST parton distribution
sets.}
\label{fig:d0dijet}
\end{figure}

	Both experiments have reported results for the dijet mass cross 
section and 
observe agreement with each other and with NLO predictions, albeit with an 
excess at
high dijet mass consistent with that predicted by the CTEQ4HJ pdf,
 as shown in Figure~\ref{fig:d0dijet}.~\cite{jodi,468}

D0 has used 
the dijet mass cross section ratio of $|\eta|<0.5$ over $0.5<|\eta|<1.0$
to test for compositeness.
At the 95\% CL limit, a compositeness scale of 2.4 TeV can be excluded. 
~\cite{468}
Previous measurements of the dijet angular distribution provided an exclusion of
${\Lambda}<$ 2.0 TeV. 

\begin{figure}
\center
\psfig{figure=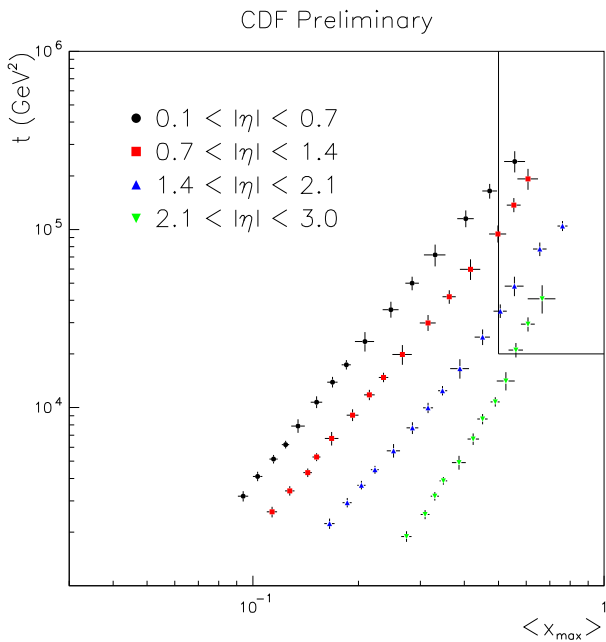,height=3.5in}
\caption{A plot of the $\hat{t}$ vs $x_{max}$ reach for the CDF differential
dijet analysis. The box in the upper right hand corner indicates  the 
kinematic region where an possible excess at HERA has been probed.}
\label{fig:tvsx}
\end{figure}

	CDF has presented a measurement of the differential dijet cross 
section in
which one jet (the trigger jet) is required to be central ($0.1<|\eta|<0.7$), 
while the other
jet (the probe jet) can have any rapidity value up to 3.0.~\cite{jodi}
 The differential 
dijet
cross section is then plotted versus the transverse energy of the trigger jet, 
for the 4
different probe jet rapidity intervals. The measurement presented in this way 
takes best
advantage of the better jet \et\ resolution CDF has in the central rapidity 
region. This
measurement also directly 
probes higher x values than the inclusive jet cross section. 
The dijet
cross section is sensitive to the high x gluon distribution and anything 
unusual that
may occur at high x and $Q^2$. This can be seen from Figure~\ref{fig:tvsx}
where $\hat{t} (=2E_T^2 cosh^2(\eta^*) (1-tanh(\eta^*))$ is plotted
versus $x_{max}$ for the dijet cross section bins. 
The box in the upper right-hand corner
indicates the region of phase space where a possible excess at HERA has been 
probed.~\footnote{For those of you who still do not get the Prince joke,
please send email to me at the address given.}

\begin{figure}
\center
\psfig{figure=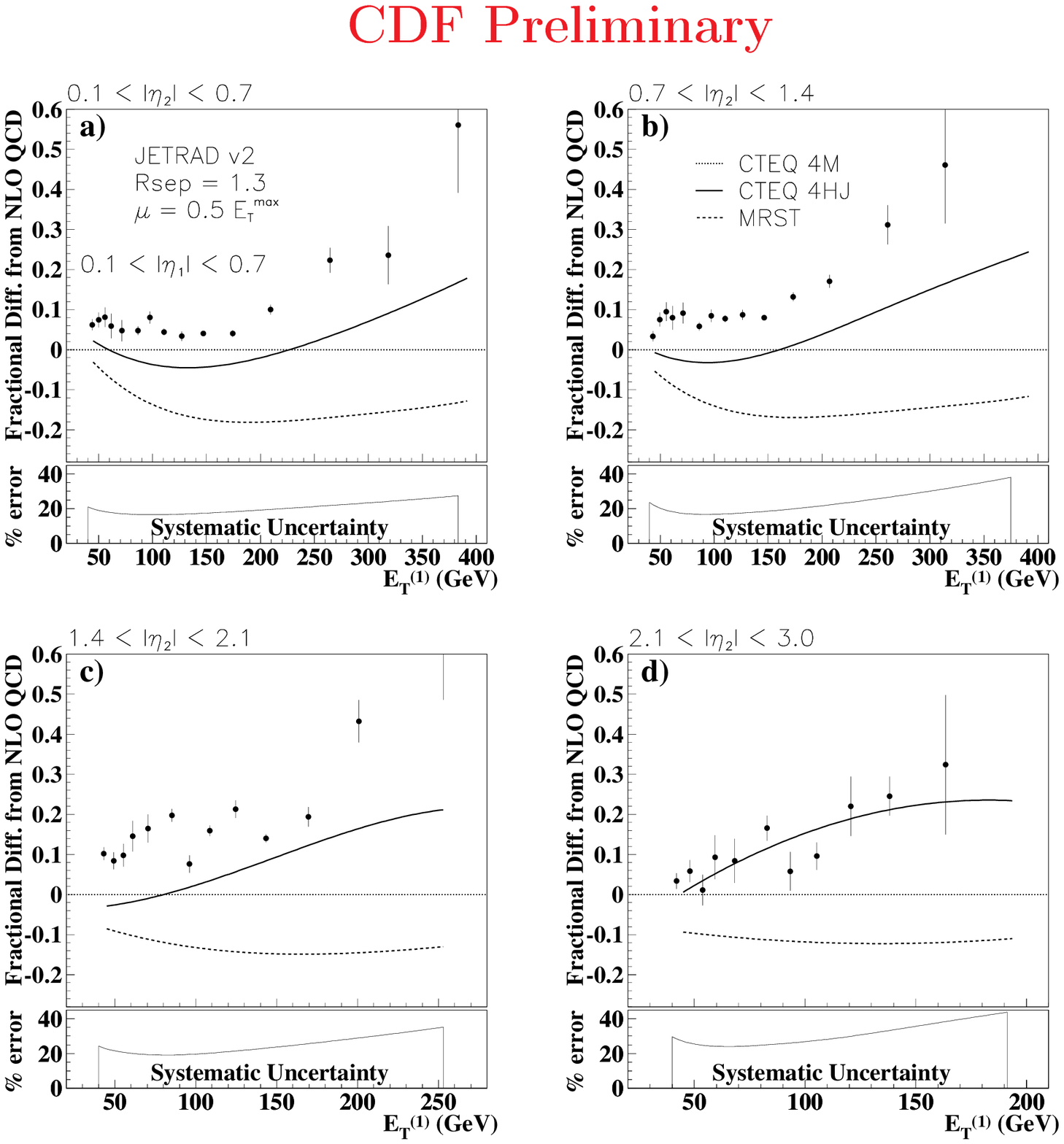,height=3.75in}
\caption{A comparison of the CDF differential dijet cross section to NLO QCD
(JETRAD)
predictions using the CTEQ4M, CTEQ4HJ and MRST parton distribution sets. 
The systematic error band is indicated at the bottom for each probe jet 
rapidity interval.}
\label{fig:diffdijet}
\end{figure}

	A comparison of the measured 
dijet cross sections to predictions using the CTEQ4M,
CTEQ4HJ and MRST pdf's is shown in Figure~\ref{fig:diffdijet}.
An excess is observed at high \et\
(corresponding to high x) for each of the probe jet rapidity   bins. The size 
of the excess
is decreased when the CTEQ4HJ pdf is used. A detailed conclusion may wait, 
though,
until a more detailed study of the adequacy of the NLO predictions is carried 
out for
the high $E_T$, high $\eta$ region. (At very high x, multijet final states
are common and NLO phase space may not be adequate.) 
The MRST predictions are uniformly below the data at all
\et\ values  due to the weaker gluon distribution discussed previously. 

	Whenever two jets of roughly equal $E_T$ values are separated by a
large rapidity interval, the emission of gluons in the rapidity region between
the two jets generates logarithmic contributions 
[${({\alpha}_s\ln(s/p_T^2))}^n$] to the dijet cross section which need 
to be resummed using the BFKL equation.~\cite{orr} Naive BFKL predictions lead
to behavior that differs dramatically from that obtained from fixed-order 
perturbative predictions. The imposition of kinematic constraints on the
calculations, however, supresses the BFKL-like behavior at the Tevatron and
reduces the chances of unambiguous observation of BFKL signatures. The 
kinematic environment at the LHC will be more favorable for observation
of BFKL-like behavior.~\cite{orrstir}
 
\subsection{Jet Production at 630 GeV and the $x_T$ Scaling Ratio}

\begin{figure}[t]
\center
\psfig{figure=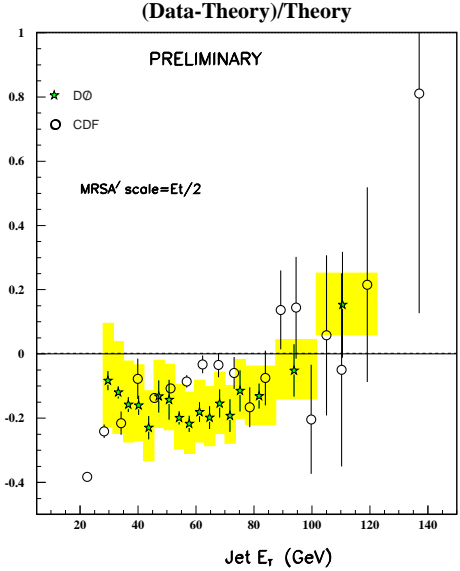,height=3.5in}
\caption{A comparison of the CDF and D0 inclusive jet cross sections at
630 GeV to NLO QCD predictions using the MRSA' parton distribution set. The
shaded band indicates the D0 systematic error.}
\label{fig:630}
\end{figure}

\begin{figure}[t]
\center
\psfig{figure=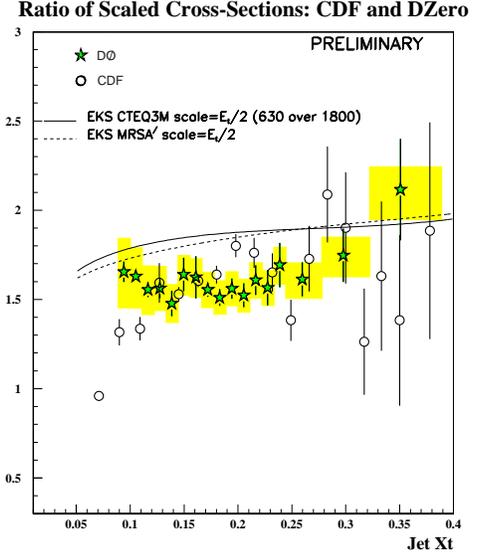,height=3.5in}
\caption{A comparison of the scaled inclusive jet cross sections (630/1800)
for CDF and D0 to NLO QCD predictions. The shaded band gives the D0 systematic
errors.}
\label{fig:cdfd0xt}
\end{figure}

	CDF and D0 have both presented  measurements of the inclusive jet 
cross section
at 630 GeV, and of the $x_T$ scaling ratio.~\cite{blazey} 
A comparison of the  
cross sections of the two experiments to
NLO predictions is shown in Figure~\ref{fig:630}. 
For both experiments, deviations from 
the NLO 
predictions are observed for jet \et\ values below 90-100 GeV/c. When the 
scaled cross
section ratio ($1/(2\pi)E_T^3d{\sigma}/dE_T$  for 630/1800) is plotted versus 
$x_T$ (=$2E_T/{\sqrt{s}}$), many of the
systematic uncertainties cancel for both experiment and theory. A naive 
parton model
prediction would give a value for the ratio of 1.0; QCD effects change the 
prediction to
closer to 2 (with some dependence on $x_T$ as observed). 
Both experiments have measured 
an $x_T$
scaling ratio lower than the theoretical prediction for $x_T<0.3$. (See
Figure~\ref{fig:cdfd0xt}.)
A similar 
discrepancy was 
observed for an earlier comparison of 546 GeV jet data to 1800 GeV jet 
data.~\cite{546} 
The reason
for the discrepancy is still under theoretical investigation and may be due 
to a 
combination of effects (underlying event subtraction, initial state \kt\, and 
additional non-perturbative jet 
fragmentation
effects (``splashout'')).~\cite{xtexplain}

\subsection{W + Jet(s) Production at the Tevatron}

	Another ratio of observables with reduced systematic errors is the 
ratio  of W + 
jet(s)
production to  W production. This measurement is naively sensitive to the
value of
$\alpha_s$ and, in fact, was originally proposed as a means of measuring 
$\alpha_s$. 
D0 has reported an 
exclusive measurement of W+1 jet production to W + 0 jet production at 
several recent
conferences. The jets were measured with the standard D0 jet algorithm 
using a cone 
radius of 0.7. The result was in serious disagreement with the NLO QCD 
predictions and
no choice of scale or pdf provided any significant improvement. 

\begin{figure}
\center
\psfig{figure=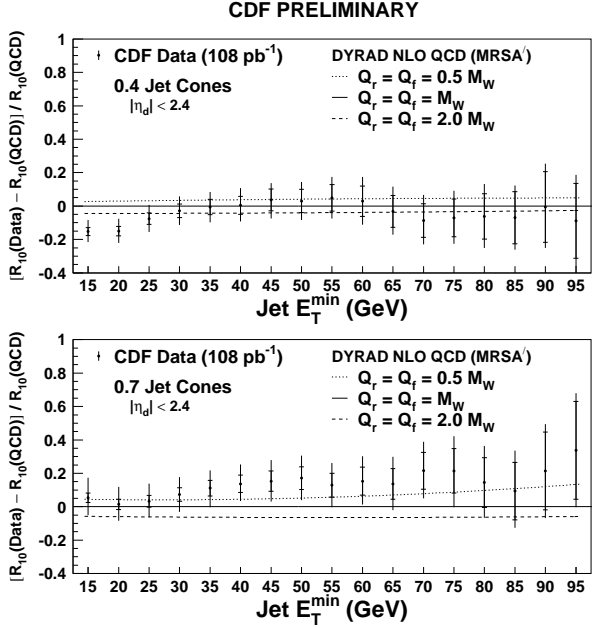,height=3.5in}
\caption{The ratio of $(R_{10}(Data)-R_{10}(QCD))$ to $R_{10}(QCD)$ for
NLO QCD predictions calculated using DYRAD. The parton distribution 
function set used is MRSA', and the baseline renormalization and 
factorization scale is $M_W$. Also shown are curves using other $Q^2$
scales.}
\label{fig:r1047}
\end{figure}

\begin{figure}
\center
\psfig{figure=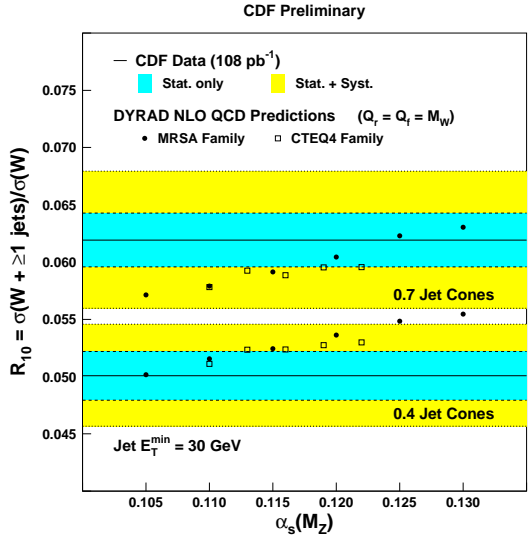,height=3.0in}
\caption{The ratio $R_{10}$ at $E_{Tmin}$ = 30 GeV for both 0.4 and 0.7 jet
cones, compared to DYRAD predictions as a function of ${\alpha}_s(M_Z)$.
The CDF data are shown as horizontal bands. The theoretical values for the
MRSA' and CTEQ4 pdf families are shown as solid circles and open squares,
respectively.} 
\label{fig:r10als}
\end{figure}

\begin{figure}
\center
\psfig{figure=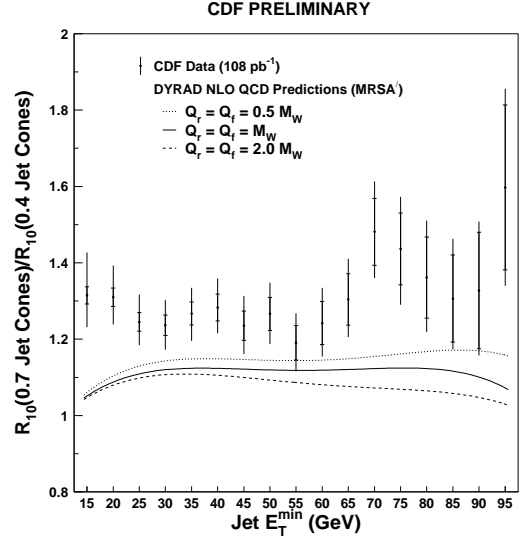,height=3.0in}
\caption{The ratio of $R_{10}$ with a jet cone size of 0.7 to $R_{10}$ with
a jet cone size of 0.4 for both CDF data and NLO QCD theory. Theory
predictions are shown for several choices of scale.}
\label{fig:r10rat74}
\end{figure}

	CDF has measured the ratio of W+$\ge$1 jet to inclusive W production 
($R_{10}$) using a jet
cone radius of 0.4~\cite{r10p4} and (new for this conference) 0.7.~\cite{r10p7}
A comparison of both
cone size results to NLO QCD predictions ~\cite{dyrad} in Figure~\ref{fig:r1047}
indicates good 
agreement. In Figure~\ref{fig:r10als},
the CDF data is compared to NLO predictions using a 
variety
of pdf's corresponding to different $\alpha_s$ values. 
The theoretical predictions have a 
surprisingly small dependence on the value of $\alpha_s$. 
The experimental ratio of $R_{10}$ with
0.7 cones to $R_{10}$ with 0.4 cones is 
shown in Figure~\ref{fig:r10rat74} along with the theoretical 
predictions. 
The experimental ratio is larger than the theoretical one, which is another 
indication that 
jets at the Tevatron are broader than the theoretical predictions. A 
subtraction of an extra
underlying event energy contribution would improve the agreement of the 
experimental 
jet shape with theory, and thus improve the agreement of the experimental 
$R_{10}$(0.7/0.4) ratio with the theoretical one. 

\section{Tevatron Fixed Target}

\subsection{Direct Photon Production}

	Direct photon production has long been viewed as an ideal vehicle for
measuring the gluon distribution in the proton.~\cite{kt1} The
quark-gluon Compton scattering subprocess ($gq{\rightarrow}{\gamma}q$) 
dominates $\gamma$ production in
all kinematic regions of pp scattering, as well as for low to moderate values
of parton momentum fraction \x\, in $\overline{p}p$ scattering. 
As mentioned previously,
the gluon distribution is relatively well constrained at low \x\ ($x<0.1$) by
DIS and DY data, but less so at larger $x$. Consequently, direct photon data
have been incorporated in several modern global parton distribution function
analyses and can, in principle, provide a major constraint on the gluon
distribution at moderate to high x.~\cite{pdfphot,CTEQ4M}

	A pattern of deviations of direct photon data from 
NLO predictions has been 
observed,~\cite{ktorig}  however, 
with the deviations being particularly striking for the E706
experiment.~\cite{e706} 
The origin of the deviations lies in the effects of initial 
state soft gluon
radiation, or $k_T$. Direct evidence of this \kt\ has long been evident from 
Drell-Yan, 
diphoton and heavy quark measurements.~\cite{ktpaper,frixione} 
The values of \avkt\ per parton vary from ~1 GeV/c at
fixed target energies to 3-4 GeV/c at the Tevatron collider. The growth is 
approximately logarithmic with center of mass energy. 

\begin{figure}[t]
\center
\psfig{figure=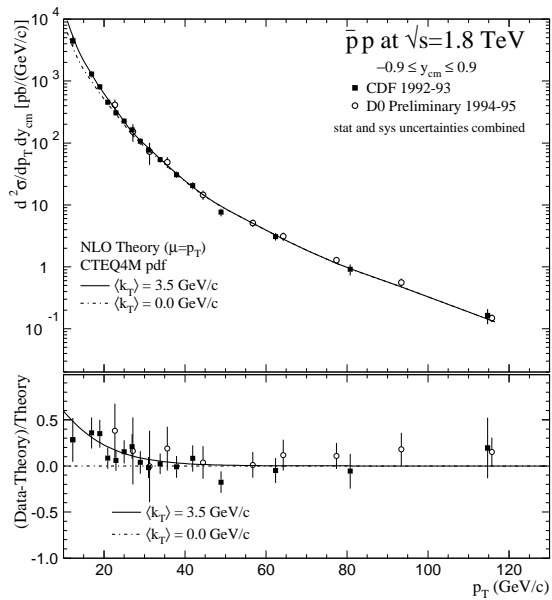,height=3.5in}
\caption{The CDF and D0 isolated direct photon cross sections, compared
to NLO QCD theory without \kt\ (dashed) and with \kt\ enhancement for
\avkt\ = 3.5 GeV/c (solid), as a function of $p_T$.}
\label{fig:tevphot}
\end{figure}

	Perturbative QCD corrections at the NLO level are insufficient to 
explain
the size of the observed \kt\, and full resummation calculations are required 
to explain
Drell-Yan, W/Z and diphoton distributions.~\cite{resum} 
These resummation calculations
correctly describe the growth of the \avkt\ with center-of-mass energy.
In a \kt\ type of resummation 
calculation,
there are typically two scales involved, which are important to the problem, 
and which
differ greatly in magnitude from each other. In the case of direct photon 
production, the
two scales are the mass and \pt\ (or \kt\ ) of the photon-jet system. Instead of 
examining 
the effects of the soft gluon resummation on the photon-jet mass cross 
section per se, 
one can instead look at the effects on the \pt\ distribution of the 
photon alone.

	Currently, there is no rigorous \kt\ resummation calculation available 
for single
photon production. The calculation is quite challenging in that the final 
state parton 
takes part in soft gluon emission and in color exchange with the initial 
state partons,
in contrast with the Drell-Yan and diphoton cases.
Also, the calculation is complicated by the fact that several overlapping 
power-suppressed contributions can contribute.  In lieu of a rigorous 
calculation of the
resummed direct photon \pt\ distribution, the effects of soft gluon radiation 
can be
approximated by a convolution of the NLO cross section with a Gaussian \kt\ 
smearing
function.~\cite{ktpaper,jeff} The value of \avkt\ to 
be used for each kinematic regime is 
taken directly from relevant experimental observables, rather than from a
theoretical prediction. 

	The behavior of the \kt\  smearing correction is quite different for the
Tevatron collider and for fixed target energies. In Figure~\ref{fig:tevphot}
is shown the 
comparison of NLO theory calculations (with and without the \kt\ corrections) 
to the direct photon data from CDF and D0. The value of \avkt\ used (3.5 GeV/c)
was taken directly from diphoton measurements at the Tevatron.~\cite{linn}

	There are two points to note: (1) the agreement with the data is 
improved if the \kt\ correction is taken into account and (2) the \kt\ smearing
effects fall off roughly as $1/p_T^2$. The latter behavior 
is the expectation for such a power-suppressed type of effect. 

\begin{figure}[t]
\center
\psfig{figure=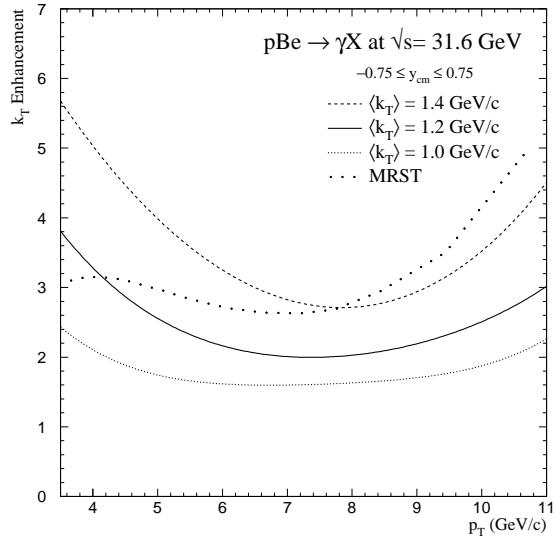,height=3.5in}
\caption{The variation of $k_T$ enhancements 
(ratio of cross sections with and without the $k_T$ corrections)
relevant to E706 direct photon
data at 31.6 GeV, for different values of average $k_T$ . 
Also shown is the
$k_T$ correction for E706 used in the recent MRST fit.}
\label{fig:kfac}
\end{figure}

	The \kt\ correction obtained for E706 at a center of mass energy of 31.6
GeV is shown in Figure~\ref{fig:kfac}. 
The value of \avkt\ of 1.2 GeV was obtained from 
measurements of several kinematic variables in the experiment.~\cite{ktpaper}
The
\kt\ smearing effect is much larger here than observed at the collider and does 
not have the $1/p_T^2$ falloff. 
This can be understood from the following argument.
At low \pt\, an \avkt\ value of 1.2 GeV/c is non-neglible compared to the \pt\
in the hard scattering, and the addition of the \kt\ smearing both increases the
size of the cross section and steepens the slope. At high \pt\ (corresponding
to large $x$), the unmodified NLO cross section becomes increasingly steep 
(due to the rapid falloff of the parton densities) and hence the effect
of the smearing again becomes larger. 

\begin{figure}[t]
\center
\psfig{figure=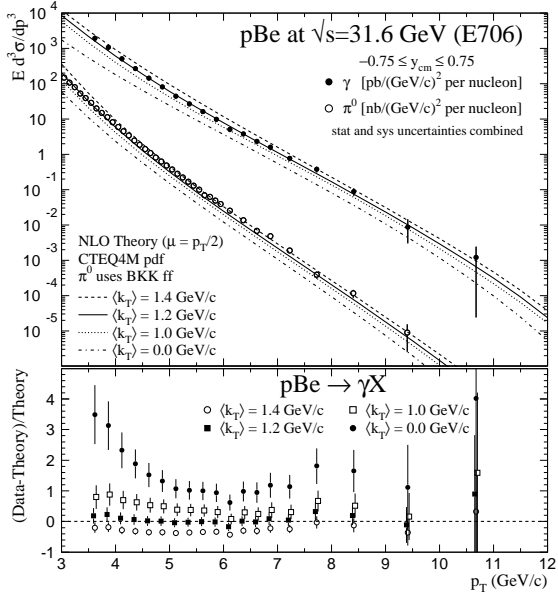,height=3.5in}
\caption{The photon and ${\pi}^o$ cross sections from E706 compared to
$k_T$-enhanced NLO QCD calculations. In the bottom plot, the quantity 
(Data-Theory)/Theory is plotted, using $k_T$-enhanced calculations for
several values of $<k_T>$. The error bars have experimental statistical
and systematic errors added in quadrature.}
\label{fig:5304m}
\end{figure}

	For both the fixed target and collider cases, the \kt\ smearing increases
the size of the observed cross section. The direct photons are only measured
above a given threshold. The direct photon \pt\ distribution
 is steep and the net
effect of the \kt\ smearing (or a more complete resummation treatment) is to
transport events from below the threshold to above the threshold, thus 
increasing the observed cross section. 

	The uncertainty in the value of \avkt\ is estimated by the E706 authors
~\cite{ktpaper} to be ${\pm}0.2$ GeV. The effect of this variation on the \kt\ 
correction is 
shown in Figure~\ref{fig:kfac} and can be observed to be quite sizeable. 
Also shown is the
$k_T$ correction used in the recent MRST paper.~\cite{mrst} The MRST paper 
uses an
analytic \kt\ smearing correction technique with an \avkt\ per 
parton value of 1.3 GeV/c.~\footnote{The MRST paper quotes a 
smaller value for the \avkt\ used for
E706. This smaller value is equivalent to 1.3 GeV/c in the convention
used in this talk and in Reference 29.}

	The E706 direct photon cross sections for pBe collisions at \s\ = 31.6 
GeV is shown in Figure~\ref{fig:5304m}
along with the NLO theoretical predictions for the
range of \kt\ corrections.~\cite{e706,ktpaper,carl} 
Very good agreement is obtained with the use of
the nominal value of \avkt\ ; the experimental cross section differs both in
magnitude and shape from the NLO prediction with no \kt\ correction.
Also shown in the figure are the data and theoretical predictions for 
${\pi}^o$ production. NLO calculations for ${\pi}^o$ production have a 
greater uncertainty than those for direct photon production since they
involve parton fragmentation. However, the \kt\ effects are expected to
be generally similar to those observed in direct photon production, and the
${\pi}^o$ data can be used to extend tests of the consequences of \kt\
smearing. The presence of an additional \kt\ value of a magnitude similar 
to that needed for single photon production leads to a substantially 
improved agreement between the ${\pi}^o$ data and theory.

\begin{figure}[t]
\center
\psfig{figure=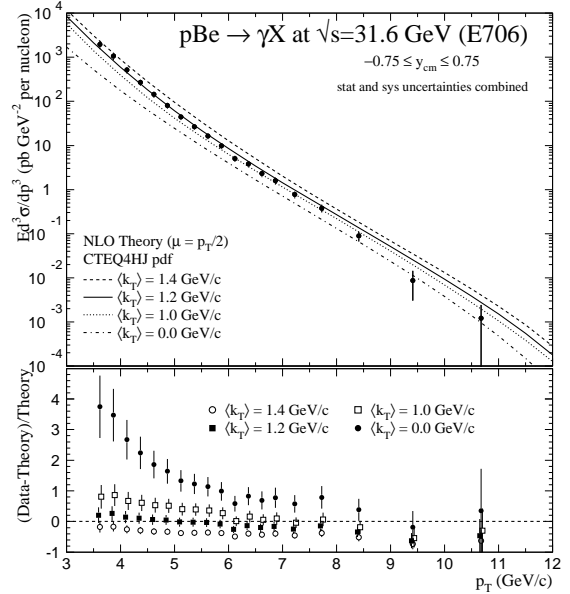,height=3.5in}
\caption{The photon cross section from E706 compared to $k_T$-enhanced
NLO QCD calculations using the CTEQ4HJ parton distribution set.}
\label{fig:5304hj}
\end{figure}

	The
same comparison is made in Figure~\ref{fig:5304hj}
using the CTEQ4HJ pdf's. Good agreement
is observed at low \pt\, but the theoretical prediction is larger than the
data at high $p_T$. As mentioned previously, the dominant mechanism for direct
photon production is gluon-quark scattering, so an increase in the gluon
distribution in this range by a factor of 2 leads to an increase in the direct
photon cross section by the same factor. Similar conclusions are obtained with
the E706 data at \s\ = 38.8 GeV.~\cite{ktpaper,carl} 

	A decrease in the \avkt\ per parton at high \pt\ would lead to a better
agreement of the E706 data with the CTEQ4HJ pdf predictions. There are 
several possible suppression mechanisms for soft gluon emission in this
kinematic region that are not taken into account in  the simple \kt\
model discussed above.~\cite{ktpaper}
(An experimental
 determination of the dependence of the $<k_T>$ as a function of $p_T/x$
is difficult due to the diminishing statistics at higher $p_T$.)
This possibility is currently under investigation. 

\begin{figure}[t]
\center
\psfig{figure=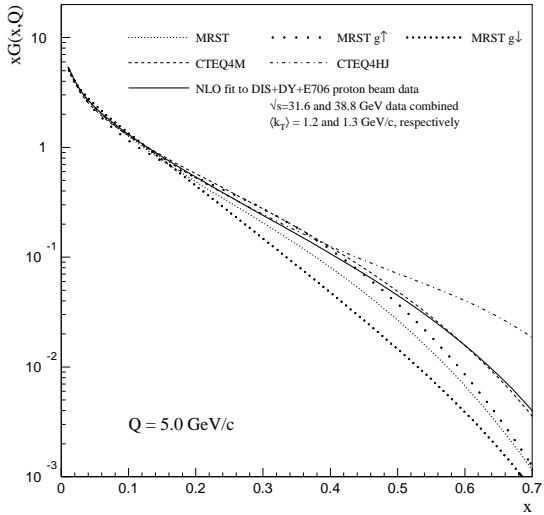,height=3.5in}
\caption{A comparison of the CTEQ4M, MRST and CTEQ4HJ gluons and the 
gluon distribution derived from fits that use E706 data. The $g\!\!\uparrow$
and $g\!\!\downarrow$ gluon densities correspond to the maximum variation in
\avkt\ that MRST allowed in their fits.}
\label{fig:gluon}
\end{figure}

	A comparison of the CTEQ4M and CTEQ4HJ gluon distributions and the
gluon obtained by fitting the \kt\-corrected E706 data (along with the 
CTEQ4 DIS and DY data sets)~\cite{ktpaper} is shown in Figure~\ref{fig:gluon}. 
As might have been expected
from Figure~\ref{fig:5304m}, the gluon distribution obtained from this fit agrees well with
the CTEQ4M gluon distribution and lies below the CTEQ4HJ gluon distribution
at high x. Also shown in the figure is the gluon distribution from the MRST
pdf, which incorporates the $k_T$-corrected data from WA70 in the
fit.~\footnote{In the MRST paper, the E706 data are not used directly in the
fits, but the fit results are compared to the data from 
E706 after 
applying an appropriate $k_T$ correction.~\cite{thankmrst} Good agreement
is obtained.} 
The larger \kt\ correction for E706 observed in Figure~\ref{fig:kfac} 
implies a smaller gluon 
distribution as is observed in Figure~\ref{fig:gluon}. 

	As has been discussed, there is a great deal of theoretical
uncertainty in the calculation of the fixed target direct photon cross
sections. Even if the Gaussian \kt\ smearing ansatz given above were formally
correct, the uncertainty in the value of \avkt\ to be used leads to a large
variation in the predicted cross section. This variation makes the use of
fixed target direct photon data in pdf fits somewhat problematic. A more 
complete resummation calculation should, with the appropriate experimental
input, be able to predict the \avkt\ per parton for each kinematic condition and
thus may be able to rescue the situation. 

	There has also been much recent interest in studying the effects of
resumming large logarithms of the form $\ln(1-x_T)$.~\cite{sudakov} 
As $x_T$ approaches 1, for any
hard scattering process, the perturbative cross section is enhanced by
powers of $\ln(1-x_T)$ that have to be resummed at all orders. These types of
effects should currently be negligible for direct photon production at the
Tevatron collider (because data are only available for relatively small values
of $x_T$), but may be important at fixed target energies. The net effect for
E706 is a significant increase in the cross section at high 
$p_T$.~\cite{mangano} 
A treatment that
includes both $k_T$ and threshold resummation effects may be necessary for a
more satisfactory description of the fixed target data. Recent theoretical
progress has been made in this direction.~\cite{li}

	In the CTEQ4 pdf fits (and the upcoming CTEQ5 fits as well), the
inclusive jet cross sections from CDF and D0 provide an additional constraint
on the gluon distribution at moderate to large x values. Because of the
theoretical uncertainties mentioned above, the CTEQ5 fits will not use 
fixed target direct photon data. 

\section{HERA}

	HERA is a positron-proton collider (27.5 GeV $e^+$ on 820 GeV protons)
with a total center of mass energy of about 300 GeV. The large center-of-mass
energy available at HERA offers a large phase space for the hadronic final state
in DIS events, thus allowing clean jet structures to be observed. As HERA
has continued its successful operation, the data available for analysis by
both of its experiments, H1 and ZEUS, has increased steadily. Along with
the increase in statistics has come an increase in the level of understanding
of the detector systematics, allowing for more precise comparisons of data to
perturbative QCD. 

\begin{figure}[t]
\center
\psfig{figure=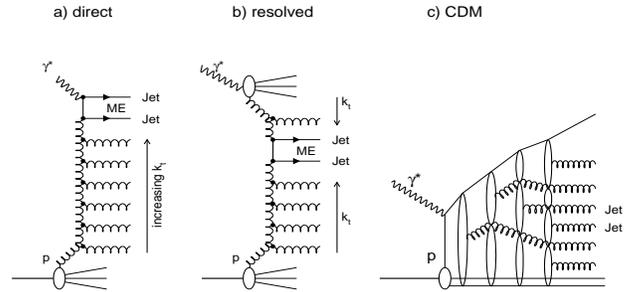,height=1.5in}
\caption{Diagrams indicating initial parton emission in ep scattering.}
\label{fig:523fig2}
\end{figure}

\subsection{Parton Evolution Dynamics}
	
	Of particular interest at HERA are measurements which discriminate
among parton evolution schemes. DGLAP~\cite{dglapiain} and BFKL~\cite{bfkliain}
describe the evolution towards large values of $Q^2$ and 
$1/x_{Bj}$, respectively. DGLAP evolution resums terms of the form 
$\ln (Q^2/Q_o^2)$ and
involves a strong ordering in
$k_T$ of the gluon emissions with the hardest emissions occurring near the 
top of the gluon ladder. (See Figure~\ref{fig:523fig2}a.) 
In the BFKL model, terms of the form $\ln(1/x)$ are
resummed and
gluon emissions are not ordered in transverse momentum $k_T$. 
 A solution of the parton evolution equation by CCFM~\cite{ccfm}
approximates the BFKL equation in the low $x_{Bj}$ limit and the DGLAP 
equation in the high $x_{Bj}$ limit.
A  
lack of ordering similar to that found in BFKL evolution
 has been incorporated in the color dipole model. 
In the color dipole model, gluon emission originates from a color 
dipole that is stretched between the scattered quark and proton remnant. 
 The result is a cascade of independently
radiating dipoles with the gluons not ordered in $k_T$.
(See Figure~\ref{fig:523fig2}c.)
In addition, for processes where the photon structure is resolved, the
hardest emissions given in the QCD matrix element may occur anywhere in the
ladder, with increasingly soft emissions along the ladder towards both the
proton and the photon. (See Figure~\ref{fig:523fig2}b.) 
The results presented previously for the Tevatron collider and fixed
target measurements are all governed by DGLAP kinematics. At HERA, results
have been obtained in all of the kinematic regimes discussed above.

	A wide variety of 
leading order (in ${\alpha}_s$) theoretical programs are available for comparison
to the HERA data incorporating the above evolution schemes:

\begin{itemize}

\item BFKL calculations at the parton level and with fragmentation functions
~\cite{bfkl}
\item the DGLAP-based parton shower Monte Carlo model LEPTO~\cite{LEPTO}
\item the CCFM-based Linked Dipole Chain model LDC~\cite{LDC}
\item the color dipole Monte Carlo model ARIADNE (which assumes a chain of
independently radiating dipoles spanned by color-connected partons~\cite{ARIADNE}
\item the resolved photon model RAPGAP~\cite{RAPGAP}

\end{itemize}
In addition, there are a number of DGLAP-based calculations in next-to-leading
order of ${\alpha}_s$, including DISENT~\cite{DISENT}, MEPJET~\cite{MEPJET},
DISASTER++~\cite{DISASTER} and JETVIP~\cite{JETVIP}, that are available in
the form of flexible Monte Carlo generators.

	Comparisons between the data and the theory can be made either at
the detector level (with an appropriate Monte Carlo simulation of the detector
response), the hadron level (with corrections for acceptance and migration)
or at the parton level (correcting for hadronization effects). 

\begin{figure}[t]
\center
\psfig{figure=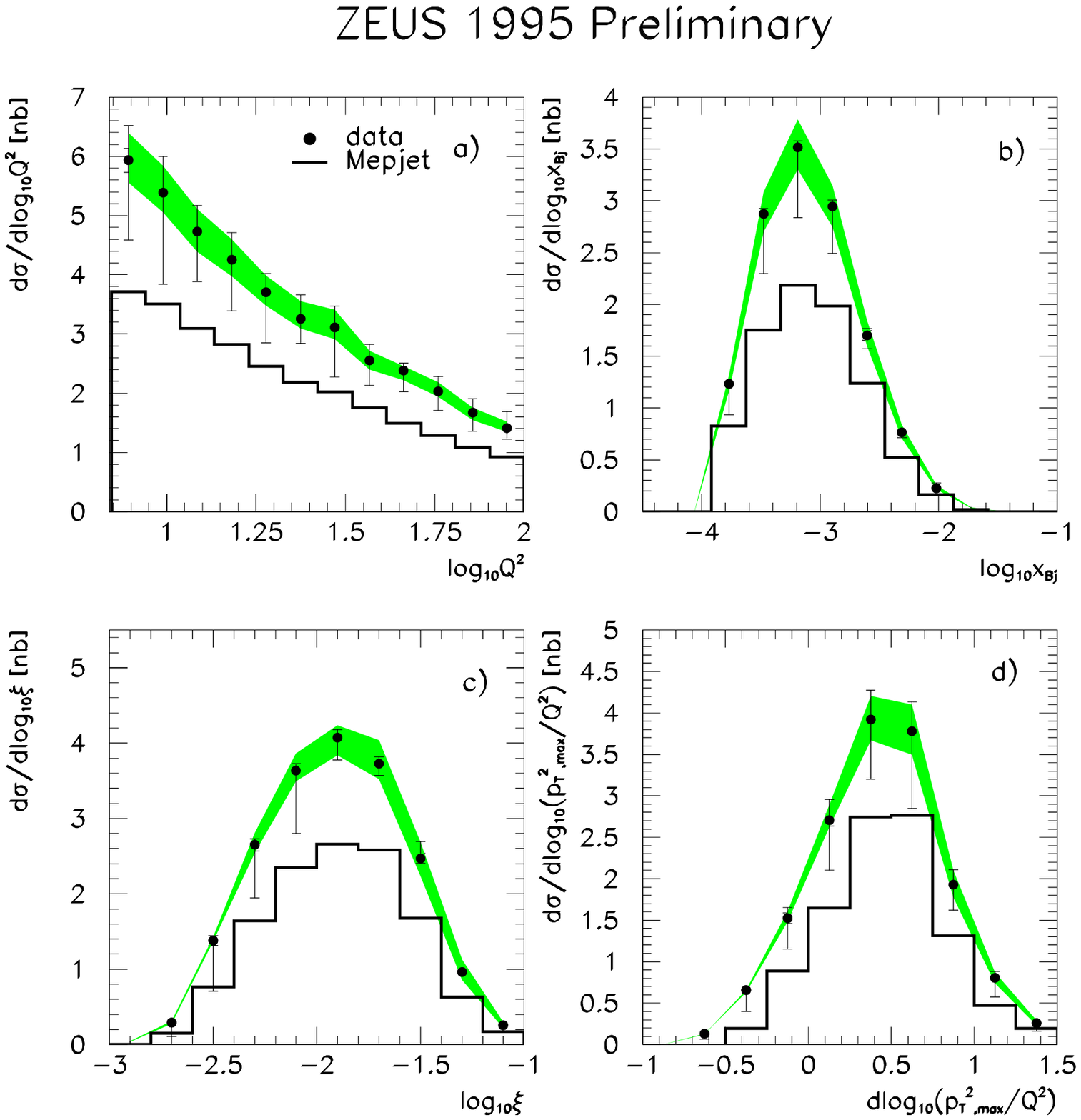,height=3.5in}
\caption{The differential dijet cross sections from ZEUS at the parton
level versus (a) $Q^2$, (b) $x_{Bj}$, (c) ${\xi}$ and (d) $p_{Tmax}^2/Q^2$.
The inner(outer) error bars indicate the statistical error (statistical
and systematic errors added in quadrature). The shaded area represents the
error due to  the uncertainty on the jet energy scale of $\pm$3.5\%. 
The solid line indicates the NLO QCD predictions from MEPJET.}
\label{fig:zeus8062}
\end{figure}

\subsection{Dijet Cross Sections in DIS}

	In DIS (in the naive quark-parton model), the virtual photon is
absorbed by a single quark or antiquark in the proton. This results in one
jet from the struck quark(antiquark) and one jet from the proton remnant
(1+1 configuration). 
To first order in ${\alpha}_s$ (lowest order for dijet production),
two jets (in addition to the proton remnant jet) 
with balanced transverse momentum are produced in the photon-proton
center-of-mass(2+1 configuration).
There are two subprocesses
responsible for dijet production in DIS: boson-gluon fusion 
(${\gamma^*}g{\rightarrow}{\overline{q}}q$) and
QCD Compton scattering (${\gamma}^*q{\rightarrow}gq$). At low \x\ and $Q^2$, 
the large gluon density
leads to the dominance of the boson-gluon fusion subprocess and allows for
a direct sensitivity for the gluon distribution in an \x\ region below the 
fixed target direct photon experiments. The presence of the strong interaction
vertex may allow a measurement of $\alpha_s$. Jet measurements have been 
conducted with variations of both the iterative 
cone and $k_T$ algorithms, and in the
lab, Breit and hadron center-of-mass reference frames.

\begin{figure}[t]
\center
\psfig{figure=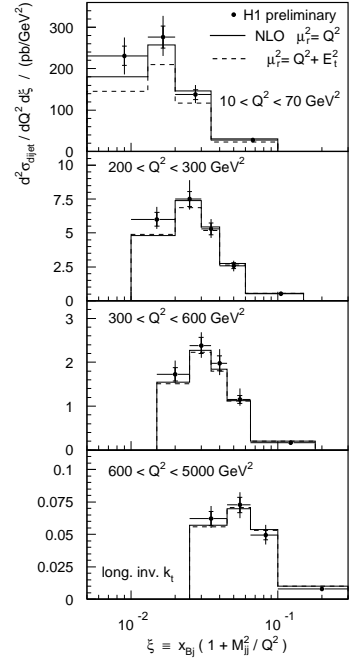,height=3.5in}
\caption{The dijet cross sections from H1 double differential in 
$Q^2$ and ${\xi}$. Also shown are the NLO QCD predictions of DISENT.}
\label{fig:h15205a}
\end{figure}

	ZEUS has submitted to this conference a measurement of dijet 
production in DIS events from a data sample corresponding to an integrated
luminosity of 6.4 $pb^{-1}$.~\cite{wobisch,repond,806}
Cuts were placed on $Q^2$(7-100 $GeV^2$), 
y($>0.04$) and
the scattered positron energy ($>10$ GeV). The jets were reconstructed with
a cone algorithm (R=1) and were required to have a transverse energy
greater than 4 GeV/c in both the laboratory and center-of-mass frames, and
to have a pseudorapidity in the laboratory frame in the range from -2 to +2.   
The differential cross sections, corrected to the 
parton level, are shown in Figure~\ref{fig:zeus8062}. 
The corrections applied are typically
20-40\%. The variable ${\xi}(=x_{Bj}(1+m_{jj}^2/Q^2)$ is of particular 
interest since it is related to the momentum fraction of the quark or gluon
emitted from the proton in the leading order QCD picture. 

	As can be observed in Figure~\ref{fig:zeus8062}, 
the data span the range from
$0.2<p_T^2/Q^2<30$, 
so the events are characterized by two scales, $p_T^2$ (square of jet 
transverse momentum) and $Q^2$. 
A comparison of an exact NLO QCD calculation (MEPJET) to the data demonstrates
that the theory adequately describes the shape of all of the distributions
but has a normalization off by approximately 34\%. The difference is comparable
in size and of the same sign as the hadronization correction applied to
convert the data to the parton level. 

\begin{figure}
\center
\psfig{figure=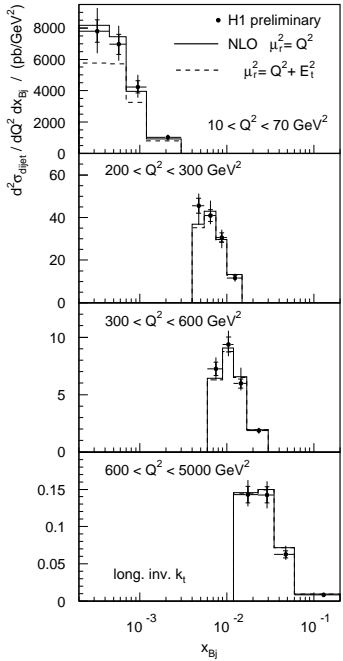,height=3.5in}
\caption{The dijet cross sections from H1 double differential in 
$Q^2$ and $x_{Bj}$. Also shown are the NLO QCD predictions of DISENT.}
\label{fig:h15205b}
\end{figure}

	The difference between theory and experiment observed by ZEUS
might originate in the
effects of soft gluon radiation in the regime (same minimum transverse 
momentum cut on both of the jets) where such effects may be important. It
has been proposed to measure the dijet rate using asymmetric cuts on the 
transverse momenta of the two jets in order to minimize this type of
correction to the NLO calculation.~\cite{kramer} Placing an additional cut on
the larger of the transverse momenta of the two jets decreases the magnitude of
both the predicted and measured dijet cross sections, but the relative 
difference between the two still remains substantial. Similarly, increasing
the cuts on the transverse momenta of the two jets results in both the
measured and calculated cross sections decreasing, with the ratio of the two
remaining approximately constant.~\cite{806} 

	H1 has measured the dijet cross section in DIS events at HERA and from
this measurement has extracted a determination of the gluon distribution in
the \x\ range from .01 to 0.1.~\cite{wobisch,repond,520} The analysis
involves a large data sample (36 $pb^{-1}$) and utilizes $k_T$ jet algorithms
for jet determination in the Breit frame. The Breit frame, where the virtual
photon collides head-on with the scattered quark, is well-suited for studies of 
dijet production. In this frame, the jet transverse energy directly reflects
the hardness of the underlying QCD process. Three different variations of the
$k_T$ algorithm were utilized with the smallest hadronization corrections 
being present in the longitudinally invariant $k_T$ algorithm. The 
double-differential dijet cross sections 
(corrected to the hadron level) are shown in Figure~\ref{fig:h15205a}
as a function
of $Q^2$ and ${\xi}$ and in Figure~\ref{fig:h15205b}
as a function of $Q^2$ and $x_{Bj}$, compared
to the NLO program DISENT. Good agreement is observed but the data have not
been corrected for hadronization effects (which, however, are expected to be
small). 

	Most of the data presented above by H1 are at higher $Q^2$ than the
previously discussed ZEUS measurement. There is some overlap, though, and
the reason for the agreement by H1 with NLO theory and the disagreement by
ZEUS is still under study. It may be related to the two studies being carried 
out
in different frames of reference, with different jet algorithms and the
jet corrections being to the
hadron level for H1 compared to the parton level for ZEUS. 

\begin{figure}
\center
\psfig{figure=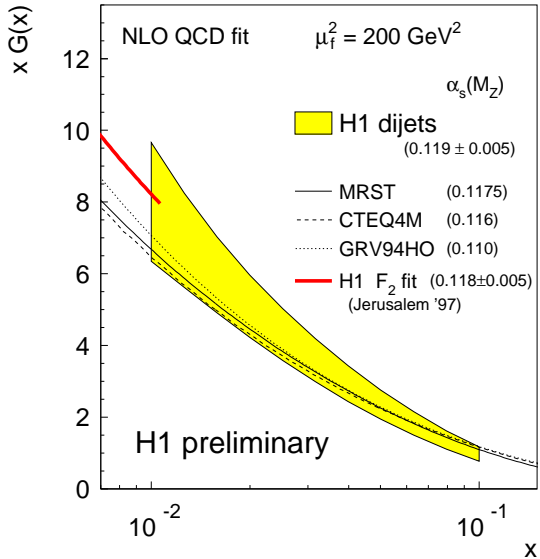,height=3.0in}
\caption{The error band of the gluon density in the proton from a NLO
QCD fit to the H1 dijet cross sections. The result is compared to the gluon
densities from different parton distribution sets and the result from a fit
to the H1 structure function data.}
\label{fig:h1gluon}
\end{figure}

	To determine the gluon distribution, the dijet cross section
distributions $d^2{\sigma}/dQ^2d{\xi}$ and $d^2{\sigma}/dQ^2dx_{Bj}$ were
utilized (in the range $200<Q^2<5000~GeV^2$, where  the NLO theory
describes the data well) along with the H1 neutral
current DIS data (to determine the quark densities in this x range). 
In the fit, the data are compared to the product of the NLO QCD prediction
and the hadronization correction. The results of the gluon fit are shown in
Figure~\ref{fig:h1gluon} 
for the \x\ range 0.01 to 0.1 with a factorization scale (${\mu}^2$)
of 200~ $GeV^2$. The result is slightly higher than the gluon distributions
obtained from several global analyses (although compatible within the error)
and is in good agreement at \x\ = 0.01 with the gluon obtained from the 
QCD analysis of the H1 $F_2$ data.

	H1 has also reported a determination of the strong coupling constant
${\alpha_s}$ from the dijet sample.~\cite{wobisch,h1alpha}
A modified JADE $k_T$ jet algorithm was
applied to a sample of NC DIS events in the $Q^2$ range from $200-10000~ GeV^2$.
The restriction of $Q^2>200~ GeV^2$ provides better acceptance for the final 
state jets and restricts the range of initial parton x to large values where 
the parton densities are better known. 

\begin{figure}
\center
\psfig{figure=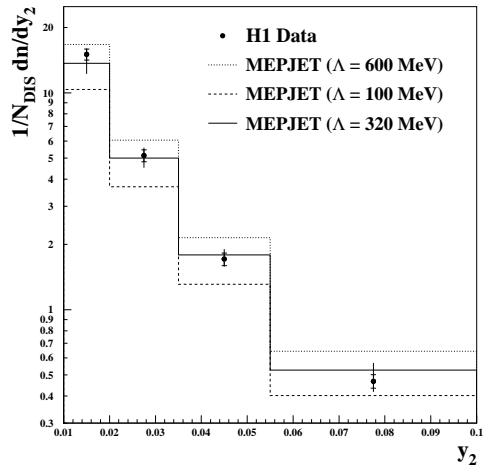,height=3.0in}
\caption{The distribution of the differential jet rate $y_2$ corrected for
detector and hadronization effects compared to the NLO prediction from
MEPJET. The full line shows the NLO prediction using the fitted value of
${\alpha}_s$.}
\label{fig:528fig5}
\end{figure}

	The jet algorithm calculates the scaled quantities $m^2_{ij}/W^2$ of
pairs of calorimeter clusters (i,j), where $W^2$ is the total invariant mass
of all clusters and $m_{ij}$ is the mass of clusters i and j. The clusters
with minimum $m_{ij}/W^2$ are added together; this procedure is repeated until
exactly (2+1) jets remain. The smallest scaled jet mass given by any
combination of the (2+1) jets is defined to be the observable $y_2$. A cut
on $y_2$ ($y_2>0.01$) is imposed to increase the fraction of events with a 
clear (2+1) jet structure, thus enhancing the sensitivity to $\alpha_s$. 
The H1 $y_2$ distribution, normalized to the number of DIS events in the
kinematic sample and corrected for detector and hadronization effects, is
shown in Figure~\ref{fig:528fig5}, along with the predictions of the NLO 
program MEPJET. Predictions are shown for ${\Lambda}_{\overline{MS}}^{(4)}$
= 100 MeV and 600 MeV, along with the best fit value of 320 MeV. This value of
$\Lambda$ corresponds to a value of ${\alpha}_s(M_Z^2)$ of 
$0.118\pm0.002^{+0.007}_{-0.018}(systematic)$. There is an additional
theory systematic error of $^{+0.007}_{-0.008}$. This result is in agreement
with the world average, albeit with a large error. The largest experimental
systematic error is due to the model dependence of the detector and
hadronization corrections; one of the largest sources of theory uncertainty
is due to the imperfect knowledge of the pdf's (and in particular the gluon
distribution) in this kinematic range.
(Gluon-initiated processes account for 
about 50\% of the (2+1) jet events used in the analysis.) There is a strong
correlation between the fit value of ${\alpha}_s$ and the size of the gluon 
distribution as was observed in determinations of ${\alpha}_s$ in jet 
production at the Tevatron.~\cite{CTEQ4M,wobisch}

\subsection{Forward Jet Production at HERA}

\begin{figure}
\center
\psfig{figure=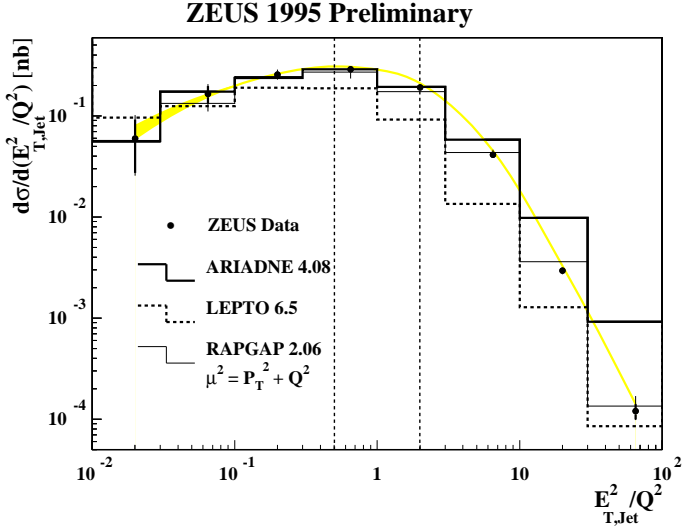,height=2.75in}
\caption{The hadron level forward jet cross section as a function of
$E_{Tjet}^2/Q^2$ from ZEUS. The data are compared to the RAPGAP Monte Carlo
model with direct and resolved contributions, to LEPTO and to ARIADNE. The 
shaded band corresponds to the uncertainty from the energy scale of the
calorimeter.}
\label{fig:zeus}
\end{figure}

	One of the significant discoveries made at HERA was the steep rise of
the proton structure function $F_2(x,Q^2)$ in the region of small 
\x\ (\x\ $<10^{-3}$). In
the BFKL approach, the leading terms in $\ln(1/x)$ 
which appear together with the
$\ln Q^2$ terms in the evolution equation are resummed. The BFKL terms may lead
to a steeper $F_2$ behavior, but from the existing $F_2$ data, 
it is not possible
to unambiguously determine whether the BFKL mechanism plays a role in the HERA
\x\ range. The BKFL mechanism predicts additional contributions to the hadronic
final state from high transverse momentum partons travelling forward in the
HERA frame. These forward-going partons may be detected experimentally as jets
and may result in an enhancement of the forward jet cross section when 
compared to either exact NLO QCD calculations or parton shower model 
calculations based on DGLAP evolution. This is the same type of BFKL physics
that was discussed earlier for large dijet rapidity separation at the Tevatron.
In this case, the large rapidity separation is between the current and forward
jets.

\begin{figure}[t]
\center
\psfig{figure=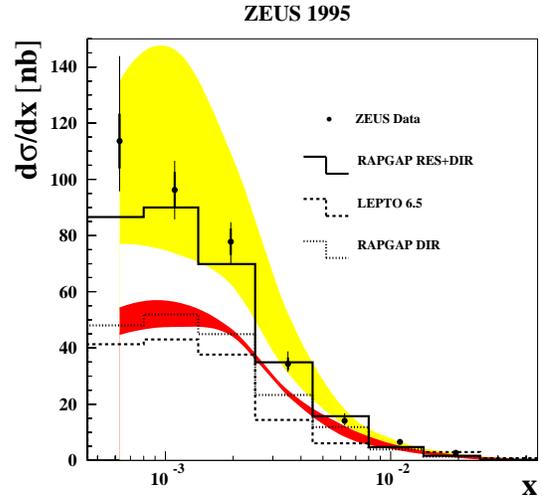,height=3.5in}
\caption{The hadron level forward jet cross section from ZEUS as a function
of $x_{Bj}$. The data are compared to the RAPGAP Monte Carlo model with
direct and resolved contributions and to LEPTO. The shaded band on the top
indicates the uncertainty due to factorization scale variation for the full
RAPGAP prediction while the shaded band on the bottom indicates the same
uncertainty for the direct contribution alone.}
\label{fig:zeusforward}
\end{figure}

	As previously, 
there are two hard scales relevant for low \x\ forward jet production:
the squared momentum transfer of the photon, $Q^2$, and the squared transverse
energy of the jet $E_{Tjet}^2$. The variable $E_{Tjet}^2/Q^2$ 
can be varied from very
small values to very large values. A low value of $E_T^2/Q^2$ corresponds to the
standard DIS regime where DGLAP dynamics is dominant. In the range where 
$E_T^2/Q^2$ is approximately 1, BFKL dynamics becomes important
(and DGLAP parton evolution is suppressed)~\footnote{Specifically, BFKL 
dynamics is important in the forward jet region when $E_T^2/Q^2$ is near 
one and $x_{jet} >> x_{Bj}$.} while in the
regime where $E_T^2>Q^2$, the jet begins to probe the structure of the photon. 
The forward jet cross section from ZEUS (corrected to the hadron level)
is shown in Figure~\ref{fig:zeus}.~\cite{milstead,804} 
All three Monte Carlo programs
shown describe the data well for $E_T^2/Q^2<<1$. ARIADNE and RAPGAP work in the
BFKL region ($E_T^2=Q^2$) while only RAPGAP is successful in the regime where 
$E_T^2>>Q^2$. The RAPGAP Monte Carlo model contains resolved as well as direct
photon contributions. A resolved virtual 
photon contribution could account for the
excess of forward jets with respect to the standard DGLAP models. 
In Figure~\ref{fig:zeusforward},
the hadron level forward cross section is plotted as a function of 
Bjorken $x$. The agreement with RAPGAP (with both resolved and direct
components) is excellent. However, the amount of resolved contribution to
the forward jet cross section has a wide range of uncertainty that
makes definitive conclusions difficult. The shaded
band in Figure~\ref{fig:zeusforward}
indicates the variation in the RAPGAP prediction when the
factorizaton scale is varied from ${\mu}^2=E_{Tjet}^2/2+Q^2$ to 
${\mu}^2=4E_{Tjet}^2+Q^2$.  
Due to the large scale dependence, a  
comparison to exact NLO calculations is needed but the appropriate hadronization
corrections are large. 

\begin{figure}
\center
\psfig{figure=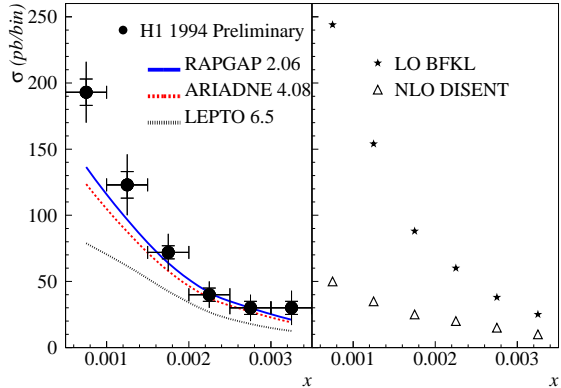,height=2.5in}
\caption{A comparison of the H1 forward jet cross section to various QCD
calculations.}
\label{fig:h1forward}
\end{figure}

	H1 has measured forward jet and dijet production requiring a 
transverse energy larger than 3.5 GeV/c in a cone of radius 
1.0.~\cite{milstead,529}
 A cut of
$0.5<E_{Tjet}^2/Q^2<2$ is applied to enhance BFKL effects. The forward jet
cross section is shown in Figure~\ref{fig:h1forward}, 
as a function of \x\, compared to several
models. ARIADNE and RAPGAP (with a resolved photon contribution) lie closer
to the data than does the DGLAP-based program LEPTO. NLO
parton calculations using DGLAP parton densities, as for example DISENT,
disagree with the data
both in shape and normalization. The numerical BFKL calculations at the 
parton level lie above the data but describe the shape fairly well. 

\begin{figure}
\center
\psfig{figure=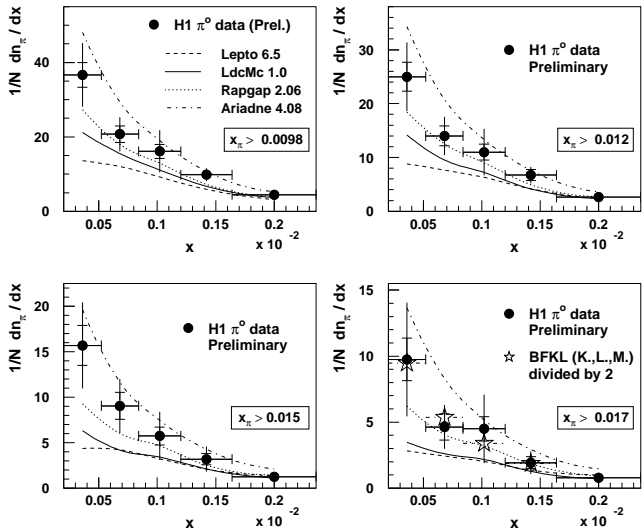,height=2.75in}
\caption{The forward ${\pi}^o$ cross sections as a function of $x_{Bj}$
for four different values of the variable $x_{\pi}=E_{\pi}/E_{proton}$
Here $n_{\pi}$ is the number of ${\pi}^o$'s and N is the number of events
in the distribution. Comparisons of four different QCD models are overlayed.
For the plot on the lower right, a BFKL calculation is shown and has been
been divided by two for presentation.}
\label{fig:h1pi0}
\end{figure}

	The forward dijet cross section is measured by H1 to be 
$6.0{\pm}0.8(stat)
{\pm}3.2(sys)~pb$, in agreement with the predictions from ARIADNE and 
RAPGAP. The 
dijet cross section is roughly 1\% of the forward jet cross section. Recent
BFKL predictions~\cite{529_20} predict on the other hand that 
3-6\% of the forward jet data should contain 2 or more forward jets. 
	
	Single high \pt\ particle production can also be used as a probe of
QCD dynamics in the forward region. The ambiguity of jet definition implicit
in the forward jet search is missing, and the smaller spatial extent allows
the measurement to smaller angles. In Figure~\ref{fig:h1pi0}, 
the forward ${\pi}^o$ data from H1
is shown plotted vs $x_{Bj}$ for several values of 
$x_{\pi}$(=$E_{\pi}/E_{proton}$).
The observed rise with decreasing $x_{Bj}$ again provides evidence
of more hard partonic radiation than predicted by DGLAP type Monte Carlo
models (such as LEPTO), and is more reasonably represented by RAPGAP, where the
photon acts as a resolved object.

\subsection{Multijet Photoproduction at HERA}

	The study of multijet photoproduction provides a direct test of
perturbative QCD predictions beyond leading order. 
Multijet kinematic
observables have been previously studied for 3-6 jet production at 
Fermilab~\cite{cdfmultijet} and for 2 jet production at HERA.~\cite{hera2jet}
ZEUS has now measured 3 jet final states in photoproduction events at 
HERA.~\cite{sinclair,800}

\begin{figure}[t]
\center
\psfig{figure=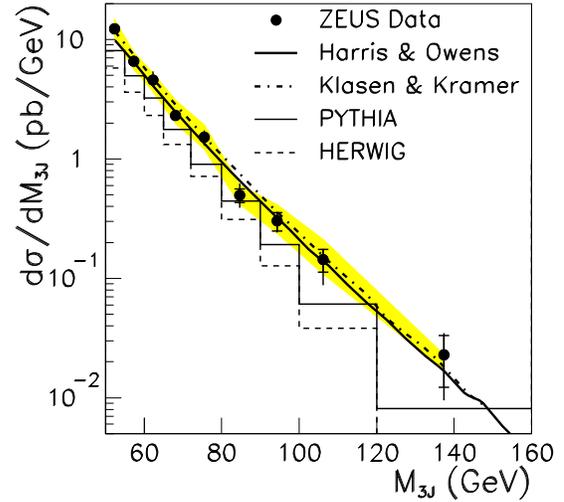,height=3.0in}
\caption{The measured three-jet cross section with respect to the three-jet
invariant mass. The inner error bar gives the statistical error and the outer
the sum of statistical and systematic errors in quadrature. }
\label{fig:800fig2}
\end{figure}

	Photoproduction events were selected by restricting the transverse
momentum to the positron to be less than 1 GeV and the photon-proton 
center-of-mass energy to be in the range from 134 to 269 GeV. Jets were defined
using a \kt\ cluster jet algorithm with the first two jets having a 
transverse energy greater than 6 GeV/c and the third a transverse energy
greater than 5 GeV/c. The requirement of relatively high transverse energy
for the jets ensures that the process can be calculated by perturbative
QCD. 

\begin{figure}
\center
\psfig{figure=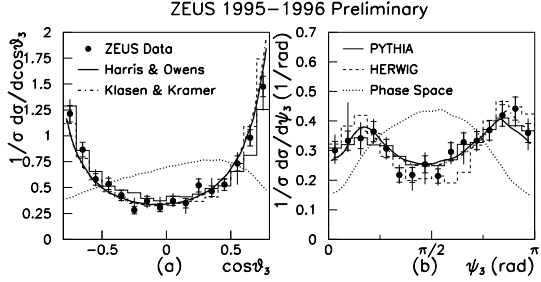,height=1.5in}
\caption{The distributions of the angles $cos({\theta}_3)$ and ${\psi}_3$.
The thick lines show the fixed order perturbative QCD calculations and the
thin lines represent the parton shower Monte Carlo predictions. The dotted 
curve shows the distribution for a constant matrix element.}
\label{fig:800fig4}
\end{figure}

	The three-jet invariant mass distribution is shown in 
Figure~\ref{fig:800fig2} and 
compared to the order ${\alpha}{\alpha}_s^2$ calculations.~\cite{harris,klasen} 
The two calculations 
are leading order for the variable under study (since there are three jets 
in the final state) but still provide good agreement with the data. The
Monte Carlo programs PYTHIA and HERWIG contain only 2 $\rightarrow$ 2 
matrix elements
but a third jet can be provided by gluon radiation. The Monte Carlo programs
predict the correct shape for the cross section but have a normalization too
low by 20-40\%.

\begin{figure}[t]
\center
\psfig{figure=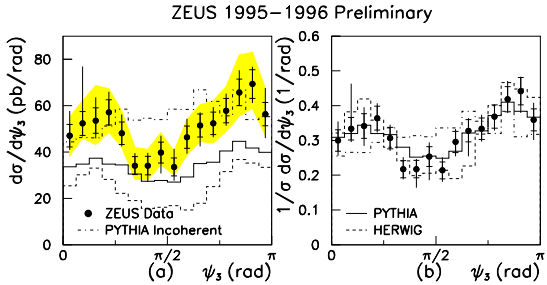,height=1.5in}
\caption{The measured cross section $d{\sigma}/d{\psi}_3$ and the 
normalized ${\psi}_3$ distributions for ZEUS. The PYTHIA and HERWIG predictions
are shown by solid and dashed lines respectively and the prediction from 
PYTHIA with color coherence switched off is shown by the dashed-dotted line.}
\label{fig:800fig6}
\end{figure}

	The distributions for ${\theta}_3$ (the angle between the highest
energy jet and the beam direction) and ${\psi}_3$ (the angle between the plane 
containing the three jets and the plane containing the highest energy jet and 
the beam direction) are shown in Figure~\ref{fig:800fig4}.

	The $cos({\theta}_3)$ distribution shows the expected Rutherford 
scattering form ($(1-cos({\theta}_3))^{-2}$). Both the ${\theta}_3$ and
${\psi}_3$ distributions differ dramatically from phase space and agree well
with both the fixed order calculations and the Monte Carlo models.

	The QCD phenomena of color coherence can be tested using the 
${\psi}_3$ distribution. In Figure~\ref{fig:800fig6} 
is shown the ${\psi}_3$ data again, 
along with predictions of PYTHIA and HERWIG (both with color coherence
implemented) and PYTHIA (with color coherence turned off). Color coherence
disfavors gluon radiation into certain angular regions which are determined
by the color flow of the primary scatter. PYTHIA with no color coherence
predicts a much flatter ${\psi}_3$ distribution than observed in either the
data or in HERWIG and (the color coherence version of) PYTHIA. 

\begin{figure}
\center
\psfig{figure=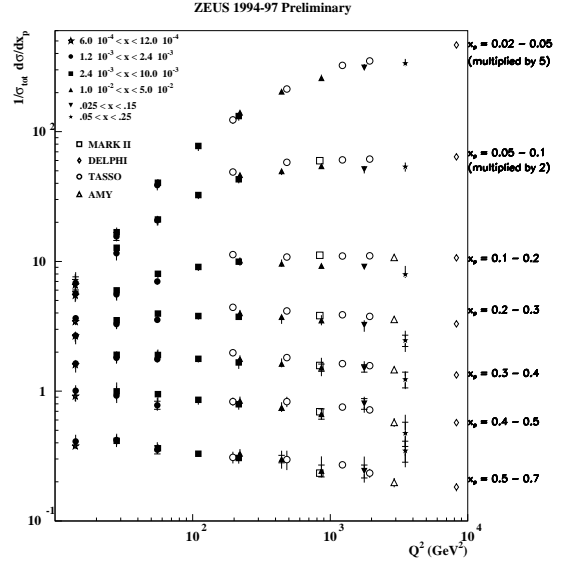,height=3.0in}
\caption{The inclusive charged particle distribution from ZEUS, in the
current fragmentation region of the Breit frame. The inner error bar is the 
statistical and the outer error bar shows the statistical and systematic
errors added in quadrature. The open points represent data from $e^+e^-$ 
experiments divided by two to take into account $q$ and $\overline{q}$
production.}
\label{fig:809fig4}
\end{figure}

\subsection{Jet Fragmentation at HERA}

	Fragmentation functions characterize the process of hadron formation
in jet production and decay. A natural frame to examine the details of jet
fragmentation in DIS events is the Breit frame, defined previously. The current
region in the Breit frame is analagous to a single hemisphere in $e^+e^-$
collisions and the fragmentation properties of these quarks can be directly
compared to the fragmentation of the struck quark in the proton. The ep
Breit frame equivalent of the $e^+e^-$ scaled hadron momentum, 
$x_p=2p_{hadron}/\sqrt{s}$, is $x_p=2p_{hadron}/Q$, where only hadrons in 
the current hemisphere are considered. The fragmentation function from the 
ZEUS experiment, plotted as a function of Q for different intervals of ${x_p}$,
is shown in Figure~\ref{fig:809fig4}.~\cite{luke,809}
Approximate scaling is observed at moderate values of $x_p$
with clear scaling violations being apparent at both large $x_p$ (decrease
with $Q^2$) and small $x_p$ (increase with $Q^2$). The HERA data overlap the
kinematic range of the $e^+e^-$ data and good agreement between both types of
experiments is observed. 

\begin{figure}
\center
\psfig{figure=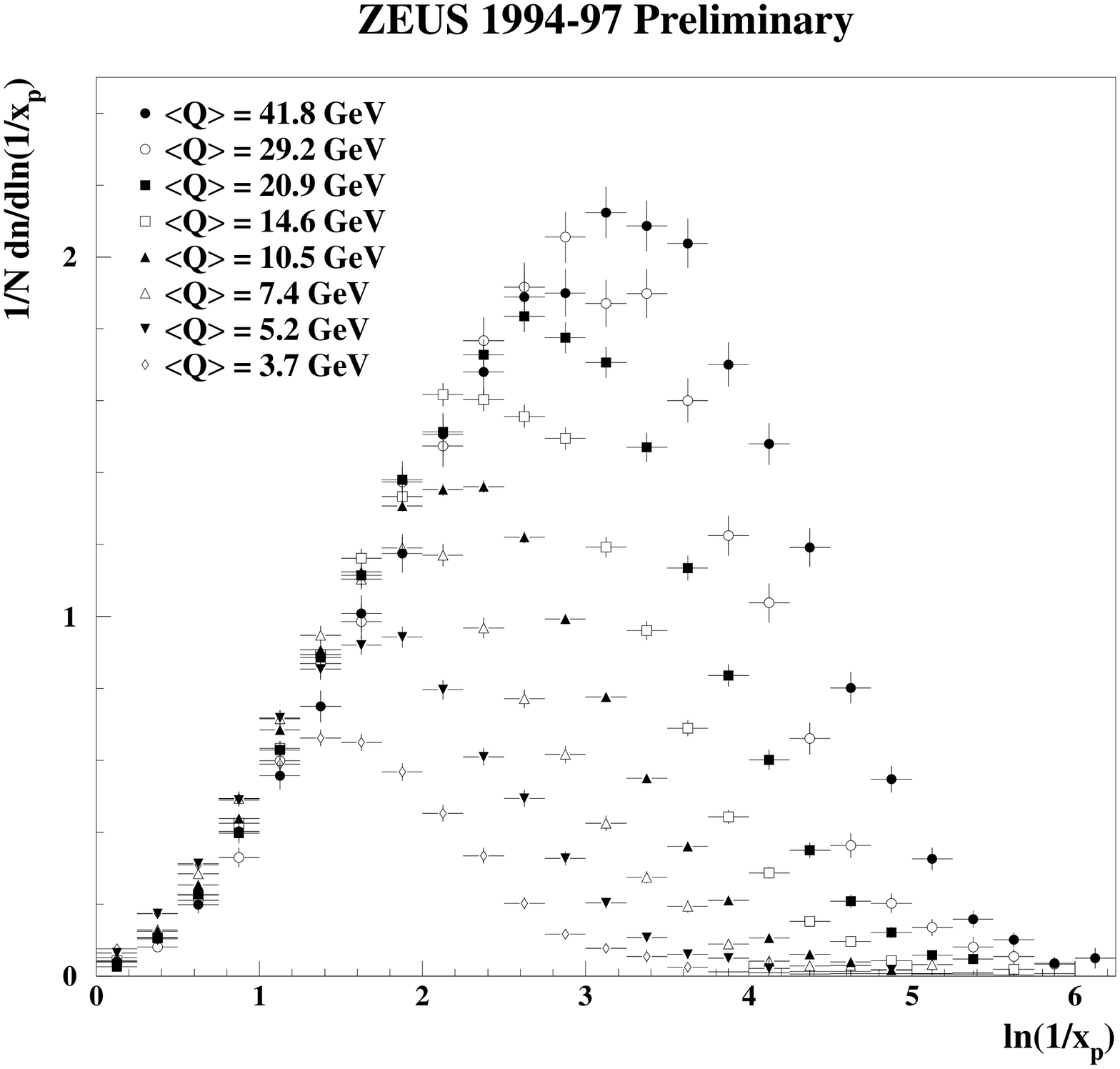,height=3.5in}
\caption{The charged particle distributions $1/N dn/d\ln(1/x_p)$ as a
function of Q. Only statistical errors are shown.}
\label{fig:809fig2}
\end{figure}

\begin{figure}
\center
\psfig{figure=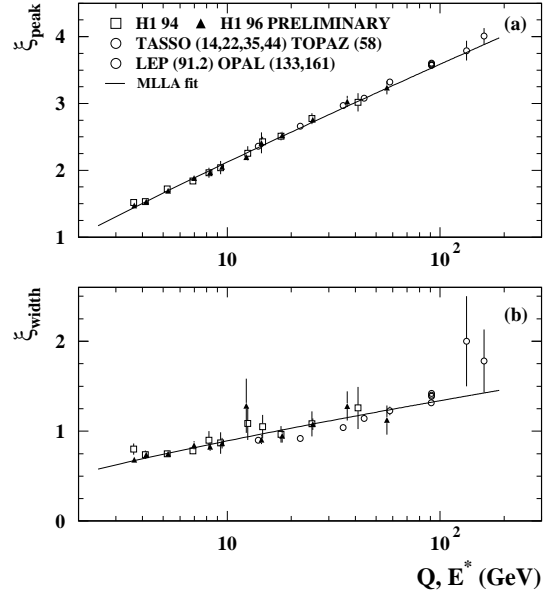,height=3.5in}
\caption{A comparison of H1 results showing the evolution of the 
peak (a) and the width (b) of the fragmentation function as a function of
Q. Also shown are $e^+e^-$ results at the corresponding values of 
center-of-mass energy. The solid line is a fit of H1 data alone to MLLA/LPHD
expectations.}
\label{fig:531fig3}
\end{figure}

	The small $x_p$ region is better investigated using the variable
${\xi}=\ln(1/x_p)$. The Modified Leading Logarithm Approximation (MLLA) together
with 
Local Parton-Hadron Duality (LPHD) predict both the shape of the $\xi$
distribution ($``hump-backed''$) and the evolution of the peak 
and the width of the ${\xi}$ distribution with energy.~\cite{MLLA}
 The charged particle
distributions are plotted as a function of ${\xi}$ in Figure~\ref{fig:809fig2}
for various Q
values, and the values of ${\xi}_{peak}$ and ${\xi}_{width}$ are shown as
a function of Q(${\sqrt{s}})$ for H1 ($e^+e^-$ experiments) 
in Figure~\ref{fig:531fig3}.~\cite{luke,531} 
The observed peak
and width of the ${\xi}$ distributions at HERA agree well with the $e^+e^-$
data and with the MLLA predictions. 

\subsection{Jet Shapes}

	For cone jet algorithms, a useful representation of the internal
structure of a jet is given by the jet shape. At sufficiently high energies,
the jet shape should be calculable in perturbative QCD, with gluon jets
broader than quark jets. At HERA, jet production has been observed in both
neutral current (NC) and charged current (CC) DIS at high 
$Q^2$.~\cite{luke,532} As mentioned previously, in DIS, jet studies can be
carried out in different frames. The appropriate frame for jet shape studies
is still under discussion and 
the effects of boosting to different frames have not
been fully investigated yet. 
 
	ZEUS has
carried out a comparison of jet shapes in NC and CC interactions at 
$Q^2>100~GeV^2$, along with jets from $e^+e^-$ and ${\overline{p}}p$ and
${\gamma}p$ collisions.~\cite{zeusshape} 
Jets are measured with an iterative cone algorithm, in the laboratory frame
of reference,  with a radius R of 1 and
an $E_T$ value of greater than 14 GeV. The jet shape was measured with the 
ZEUS calorimeter and corrected to the hadron level. 

\begin{figure}
\center
\psfig{figure=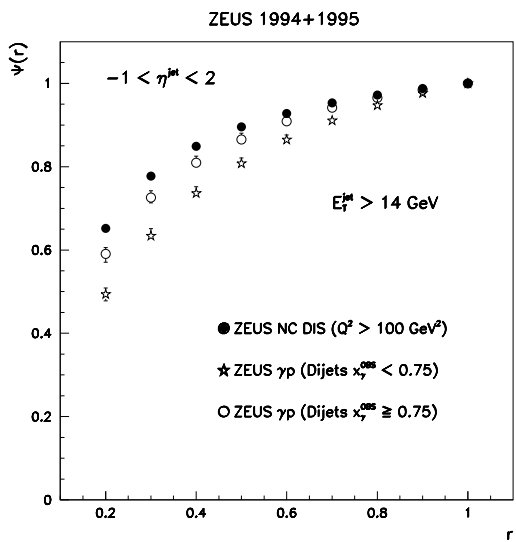,height=3.0in}
\caption{The measured integrated jet shape corrected to the hadron level from
ZEUS in NC DIS, and in resolved and direct photoproduction.}
\label{fig:shapefig7}
\end{figure}

	The integrated jet shape ${\psi}(r)$ (the fraction of jet energy
inside a cone of radius r compared to the total inside radius R)
is shown in Figure~\ref{fig:shapefig7} for 
NC events, resolved photoproduction events ($x^{obs}_{\gamma}<0.75$) and 
direct photoproduction events ($x^{obs}_{\gamma}>0.75$).~\cite{zeusshape}
In direct 
photoproduction events, the photon acts as a point-like object while in 
resolved events the parton structure of the photon is probed. 
$X^{obs}_{\gamma}$ measures the fraction of the photon's energy that goes into
the two highest $E_T$ jets. 
The jets produced
in NC DIS  are narrower than those in dijet photoproduction, but closer
to those dominated by direct processes. In NC DIS events, most of the 
final-state jets are quark jets ($e^+q{\rightarrow}e^+q$); 
direct photoproduction is
dominated by the subprocess 
${\gamma}g{\rightarrow}q{\overline{q}}$ but has contributions from 
the subprocess ${\gamma}q{\rightarrow}qg$. The resolved photoproduction events have a
larger fraction still of final-state gluon jets as evidenced by the 
larger jet width.

\begin{figure}
\center
\psfig{figure=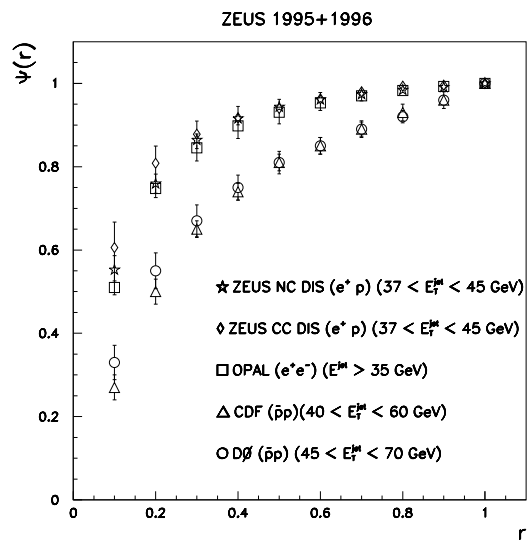,height=3.0in}
\caption{The measured integrated jet shapes corrected to the hadron level
in NC and CC DIS events at ZEUS and in $\overline{p}p$ collisions at CDF 
and D0 and from $e^+e^-$ collisions at OPAL.}
\label{fig:shapefig9}
\end{figure}

	Jets from NC and CC DIS with $Q^2>100 GeV^2$ and jet $E_T$ values
in the range from 37-45 GeV are compared to jets of similar $E_T$ values 
from CDF~\cite{jetshape6}, D0~\cite{jetshape7} 
and (from $e^+e^-$ collisions) OPAL~\cite{jetshape9} in 
Figure~\ref{fig:shapefig9}. 
For all three experiments,
an iterative cone algorithm with a radius R of 1 is used. For the two collider
experiments, an underlying event level corresponding to minimum bias events
is subtracted. No such correction is needed for the $e^+e^-$ events, or for
the HERA DIS events (in the kinematic region being considered). The jets
from the Tevatron Collider are significantly broader than the DIS jets from
HERA and the $e^+e^-$ jets from OPAL. This difference is primarily due to the
larger fraction of gluon jets expected at the Tevatron with perhaps some small
part of the difference being due to the extra contributions to the underlying
event that may be present at the Tevatron.

	H1 has carried out a measurement of the internal jet structure in
an inclusive DIS dijet sample in the kinematic domain $10 < Q^2 < 120~GeV^2$
and $2 x 10^{-4} < x_{Bj} < 8 x 10^{-3}$.~\cite{532} Jets were reconstructed
in the Breit frame using both the iterative cone and $k_T$ jet algorithms,
with the requirement that $E_{TBreit} > 5~GeV/c$ and $-1 < {\eta}_{jet,lab} 
< 2$. Jets become more collimated as $E_{TBreit}$ increases, with the 
dependence becoming more pronounced for the cone algorithm. For constant
$E_{TBreit}$, jets are narrower (broader) towards the photon (proton)
hemisphere. These dependences become smaller as $E_{TBreit}$ increases. A 
possible explanation for this behavior is that the internal jet structure
is influenced by particles close to or produced by QCD radiation near the
proton remnant. 

	Jets defined by the $k_T$ algorithm tend to be more collimated
than those defined by the cone algorithm. The dependence on $E_{TBreit}$ and
${\eta}_{Breit}$ is also stronger for the cone algorithm. Generally, the jet
shapes are well described by QCD Monte Carlo predictions with LEPTO having
a tendency to produce broader jets towards the proton remnant direction,
HERWIG producing jets which are too narrow (especially at large $E_{TBreit}$
and ${\eta}_{Breit}$), and ARIADNE lying between the two and giving a good
overall description of the data. 

\section{Conclusions}

	DGLAP-based perturbative QCD calculations have been very
successful in describing data involving jets and photons at both the
Tevatron and at HERA. Most of the areas in which the remaining 
disagreements/controversies exist involve either an uncertainty in the 
gluon distribution (the high $E_T$ jet cross section at the Tevatron) or the
influence of two scales in the measurement (fixed target direct photon
production and forward jet production at HERA). For the case of forward
jet production, BFKL effects may or may not be important; a proper
treatment of photon structure seems to describe the data both  the 
BFKL region and beyond. 
Recent calculations of the forward jet production cross sections based
on the BFKL approach showed unusually large next-to-leading order corrections,
raising the question on their predictive power.~\cite{nlobfkl} 
A deeper understanding of the
origin of these large corrections is needed before a comparison to the data
may be meaningful.

	For fixed target direct photon production
(particularly in the case of E706), soft gluon effects are extremely
important, changing both the shape and normalization of the cross section with
both the $k_T$ and Sudakov resummation formalisms being required. There has
been a great deal of theoretical effort on this problem; a successful resolution
will allow the quantitative treatment of direct photon data in pdf fits again,
provide a window on an area of very interesting physics, and finally, settle
the question of the large x gluon distribution. 

	Both CDF and D0 are undergoing major upgrades for Run 2, which is
scheduled to begin in the spring of 2000. Each experiment will put in place
a greatly improved detector and will accumulate (from 2000 to 2003) a data 
sample on the order of 2 $fb^{-1}$, a factor of 20 increase over Run 1. A data
sample this size will enable the high $E_T$ jet cross section to be probed in
much more detail as well as allowing a variety of 
Tevatron QCD measurements to be performed
with greater precision.   	

	H1 and ZEUS will continue the analysis of the data taken with 
positrons in 
1994-1997. HERA switched to electron running this year and plans to deliver 
approximately $60~pb^{-1}$ over the next two years. In 2000, the HERA machine
will be upgraded for high luminosity running, with yearly rates of $150~pb^{-1}$
expected.

\section*{Acknowledgements}
I would like to thank the organizers for a wonderful conference and a great
excuse to vacation in the Pacific Northwest. 
I would also like to thank the National Science Foundation for funding and the
following people for help in preparing this talk: L. Apanasevich, M. Begel,
J. Blazey, C. Bromberg, T. Carli, 
S. Ellis, B. Flaugher, B. Hirosky, D. Krakauer,
S. Kuhlmann, S. Linn, D. Lueke, A. Maul, S. McGill, D. Milstead, P. Nason,
J. Repond, H. Schellman, D. Soper, G. Snow, G. Sterman, J. Terron,
W.K. Tung, H. Weerts,
M. Zielinski plus all of the participants in Parallel Session 3.


\begin{thebibliography}{99}
\bibitem{thankmrst} I would like to thank James Stirling and Dick 
Roberts for supplying the numbers for the MRST \kt\ correction.
\end{thebibliography}
\end{document}